\documentclass[12pt]{extarticle}

\usepackage[a4paper]{geometry}
\usepackage{subcaption}
\usepackage{graphicx}
\usepackage{float}
\usepackage{multirow}
\usepackage{booktabs}
\usepackage{cite}
\usepackage{amsmath}
\usepackage{amssymb}
\usepackage{cancel} 
\usepackage{enumerate}
\usepackage{fancyhdr}
\usepackage{verbatim}
\usepackage{hyperref}
\usepackage{wrapfig}
\usepackage{upgreek}
\usepackage{slashed}
\usepackage{calrsfs}
\usepackage{xcolor}
\usepackage[normalem]{ulem}

\graphicspath{{Images/}}

\hypersetup{
    colorlinks=true,
    linkcolor=blue,
    }

\newcommand{\mc}[1]{\textcolor{black}{#1}}

\DeclareMathOperator*{\sumint}{
\mathchoice
    {\ooalign{$\displaystyle\sum$\cr\hidewidth$\displaystyle\int$\hidewidth\cr}}
    {\ooalign{\raisebox{.14\height}{\scalebox{.7}{$\textstyle\sum$}}\cr\hidewidth$\textstyle\int$\hidewidth\cr}}
    {\ooalign{\raisebox{.2\height}{\scalebox{.6}{$\scriptstyle\sum$}}\cr$\scriptstyle\int$\cr}}
    {\ooalign{\raisebox{.2\height}{\scalebox{.6}{$\scriptstyle\sum$}}\cr$\scriptstyle\int$\cr}}
}

\title{\mc{Higher-order-operator} corrections to phase-transition parameters in dimensional reduction}

\author{Mikael Chala, Juan Carlos Criado, Luis Gil and Javier L\'opez Miras}

\date{Departamento de F\'isica Te\'orica y del Cosmos, Universidad de Granada, Campus de Fuentenueva, E–18071 Granada, Spain\\[1cm]
\today}

\begin{document}

\maketitle

\begin{abstract}
 The dynamics of phase transitions (PT) in quantum field theories at finite temperature is most accurately described within the framework of dimensional reduction. In this framework, thermodynamic quantities are computed within the 3-dimensional effective field theory (EFT) that results from integrating out the high-temperature Matsubara modes. However, strong-enough PTs, observable in gravitational wave (GW) detectors, occur often nearby the limit of validity of the EFT, where effective operators can no longer be neglected. Here, we perform a quantitative analysis of the impact of these interactions on the determination of PT parameters. We find that they allow for strong PTs in a wider region of parameter space, and that both the peak frequency and the amplitude of the resulting GW power spectrum can change by more than one order of magnitude when they are included. As a byproduct of this work, we derive equations for computing the bounce solution in the presence of higher-derivative terms, consistently with the EFT power counting.
\end{abstract}

\newpage

\section{Introduction}
A first-order phase transition (PT) entails the sudden change on the vacuum expectation value (VEV) of a scalar field in which two non-degenerate minima coexist. Thermally-induced PTs in quantum-field theory (QFT)
are often described within the imaginary-time or Matsubara formalism~\cite{Matsubara:1955ws}, in which bosonic (fermionic) fields are periodic (antiperiodic) functions of the imaginary time compactified over a radius of size $1/T$, where $T$ is the temperature of the thermal bath. They can thus be treated as a Fourier series of thermal modes in Euclidean spacetime. In the high temperature limit, this allows for an equivalence between a 4-dimensional (4D) QFT at finite $T$ and a regular 3-dimensional (3D) Euclidean QFT.

It was long ago realised that computations within this framework present a number of difficulties, the main one being the Linde problem~\cite{Linde:1980ts}, namely the appearance of short-distance non-perturbative effects from massless vector bosons in non-Abelian gauge theories, which can be only captured with lattice simulations~\cite{Braaten:1994na}. Even in the absence of these states, there are at least two more challenges to tackle. First, loop calculations involve potentially large logarithms $\log{T^2/m^2}$, where $m$ is a light mass, which can jeopardize the validity of perturbation theory. Second, the computation of PT parameters requires the evaluation of the effective action on the so-called bounce solution~\cite{Coleman:1977py,Linde:1980tt}, that is a non-homogeneous classical field configuration that interpolates between the two VEVs. The usual construction of the effective action, built as a derivative expansion around a constant-field configuration, is doomed to fail.

These two last difficulties find an elegant solution within the realm of effective field theories (EFT)~\cite{Pich:1998xt,Manohar:2018aog}. Large logarithms can be broken into $\log{\left(T^2/\mu^2\right)}$ ---which can be minimised upon using a matching scale $\mu\sim T$---, and $\log{\left(\mu^2/m^2\right)}$ ---which can be summed using the renormalisation group equations within the EFT~\cite{Cohen:2019wxr}. A well-defined effective action, in which only the modes responsible for the thermal transition are integrated out, can be in turn constructed as an expansion in powers of $m/T$~\cite{Berges:1996ib,Strumia:1998nf,Croon:2020cgk}. In this framework, the contributions due to the non-homogeneity of field configurations such as the bounce are systematically included through effective operators containing derivatives of the field. Furthermore, the Linde problem can be also tackled this way, as the Euclidean EFT can be directly simulated on the lattice~\cite{Braaten:1994na,Farakos:1994xh}.

This is precisely the program of \textit{dimensional reduction}, the foundations of which were clarified in a seminal paper~\cite{Kajantie:1995dw} about 30 years ago (see also Ref.~\cite{Appelquist:1981vg}). Since then, it has been applied to a variety of cases~\cite{Farrar:1996cp,Cline:1996cr,Losada:1996ju,Laine:1996ms,Cline:1997bm,Laine:1997qm,Andersen:1998br,Laine:2000rm,Laine:2013raa,Brauner:2016fla,Andersen:2017ika,Gorda:2018hvi,Niemi:2018asa,Gould:2019qek,Gould:2021dzl,Ekstedt:2022bff,Gould:2023jbz,Gould:2023ovu,Niemi:2024axp,Ekstedt:2024etx}, most importantly for establishing that the PT within the Standard Model (SM) is a cross-over~\cite{DOnofrio:2015gop}. It has been also instrumental for reducing uncertainties in the determination of the gravitational wave (GW) stochastic background that ensues from strong PTs. With very few exceptions~\cite{Moore:1995jv,Croon:2020cgk,Schicho:2021gca} (see also Refs.~\cite{Chapman:1994vk,Laine:2018lgj} for calculations within thermal QCD), however, the influence of effective operators, involving more than four scalar fields and/or more than two derivatives, on this phenomenon has been neglected.

Based on a simple scalar model, in this paper we find that strong PTs occur often at parameter space points where the EFT is close to the limit of validity. This can be intuitively understood: strong PTs are characterised by $v/T\gtrsim 1$, where $v\sim m/\sqrt{\lambda}$ is the VEV after the PT and $\lambda$ stands for the scalar quartic coupling. Hence, unless $\lambda\ll 1$, $m/T$ is relative large. Within this regime, effective operators in the 3D EFT, formally suppressed by higher powers of $m/T$, can in principle not be neglected. Hence, here we study the effects of these interactions on the determination of PT parameters. On the process, we face the challenge of computing the bounce in the presence of higher-order derivative terms, which we address using perturbation theory.

The article is organised as follows. In Section~\ref{sec:theory}, we introduce the model and its corresponding dimensionally reduced EFT. We discuss the computation of the different PT parameters in Section~\ref{sec:parameters}, putting emphasis on the invariance of physical ones under (perturbative) field redefinitions. We present our main results, comparing the predictions in the presence and in the absence of effective operators, in Section~\ref{sec:results}. We conclude in Section~\ref{sec:conclusions}. Finally, we collect a number of technical details in Appendices~\ref{app:matching} and \ref{app:bounce}.

\section{Theoretical setup}
\label{sec:theory}
We consider a model consisting of a real scalar $\phi$ and a massless fermion $\psi$. The 4D Lagrangian in Minkowski space reads:
\begin{equation}
 \mathcal{L}_\text{4} = \frac{1}{2}(\partial\phi)^2 - \frac{1}{2}m^2\phi^2 - \kappa\phi^3 - \lambda\phi^4 + \overline{\psi} i\slashed{\partial} \psi - g \phi\overline{\psi}\psi\,.
 \label{eq:LagUV}
\end{equation}
At finite temperature in the imaginary-time formalism, each 4D field reduces to a tower of 3D Matsubara modes of thermal masses $m_n=2\pi n T$ ($m_n=2\pi \left(n + 1/2\right) T$) with $n\geq 0$ for bosons (fermions). Therefore, in the high-$T$ limit, the 3D EFT contains only the zero-mode of $\phi$, which we will refer to as $\varphi$.

For building the EFT, we match off-shell correlators with only the light, zeroth order Matsubara mode of $\phi$ in external legs. Since our main goal is understanding the impact of effective interactions, we
for simplicity include only the effects of modes of $\psi$ in the loops, which dominate the matching because the first non-zero mode of the fermion is lighter than that of the scalar (see Appendix~\ref{app:matching}), and because the strong PTs that can be studied within the regime of validity of the EFT necessarily have small $\lambda$ and $\kappa/T$. This implies that the EFT presents a $\mathbb{Z}_2$ symmetry $\varphi \to -\varphi$ only broken by the trilinear term.

We assume the usual power counting~\cite{Kajantie:1995dw}, $P\sim m\sim g T$, and work to order $\mathcal{O}(g^8)$. This implies that, in the EFT, the only non-vanishing interactions are those of dimension $d\leq 8$ in 4D. The most general EFT Lagrangian with (broken) $\mathbb{Z}_2$ symmetry, that we obtain with the help of \texttt{BasisGen}~\cite{Criado:2019ugp}, in Euclidean form, reads:
\begin{align}\label{eq:3deft}
    \mathcal{L}_3 &= \frac{1}{2} K_3 (\partial\varphi)^2 + \frac{1}{2}m_3^2 \varphi^2 + \kappa_3 \varphi^3 + \lambda_3 \varphi^4\nonumber\\
    &+\alpha_{61} \varphi^6 + \beta_{61} \partial^2 \varphi \partial^2 \varphi + \beta_{62} \varphi^3 \partial^2 \varphi\nonumber\\
    &+\alpha_{81} \varphi^8 + \alpha_{82} \varphi^2 \partial_\mu \partial_\nu \varphi \partial^\mu \partial^\nu \varphi + \beta_{81} \varphi \partial^6 \varphi 
        + \beta_{82} \varphi^3 \partial^4 \varphi + \beta_{83} \varphi^2 \partial^2 \varphi \partial^2 \varphi + \beta_{84} \varphi^5 \partial^2 \varphi\nonumber\\
    &+\cdots
\end{align}
The first, second and third lines of Eq.~\eqref{eq:3deft} represent the zeroth, first and second order in our perturbative expansion, respectively; the ellipses stand for higher-order terms that we neglect in our analysis. Hence, for example, $\alpha_{61}$ is of the same order as $\beta_{61}$, despite the 3D energy dimensions of the first being $[\alpha_{61}]=0$ while for the second $[\beta_{61}] = -2$.

Including EFT operators of up to $d\leq 8$ does not only allow us to explore more accurately the parameter space where very strong PT take place; most importantly, it gives us control on the validity of the EFT expansion (effects triggered by the dimension-8 interactions being significantly smaller than those of dimension-6 ones).

In the matching, we include only the dominant part of the one-loop contributions. The detailed computations can be found in Appendix~\ref{app:matching}, where we also discuss the (small) effects provided by neglected loops of scalars. Here we simply indicate the final result:
\begin{equation}
 K_3 = 1 + \frac{g^2}{12\pi^2}\,,\quad m_3^2 = m^2 + \frac{g^2 T^2}{6}\,,\quad \kappa_3 = \kappa\sqrt{T}\,,\quad \lambda_3 = \lambda T\,;
 \label{eq:dim4 matching}
\end{equation}
\begin{align}
 \alpha_{61} &= - \frac{7 \zeta(3) g^6 }{192\pi^4}\,,&
 \beta_{61} &= - \frac{7 \zeta(3) g^2}{384\pi^4 T^2}\,,&
 \beta_{62} &= \frac{35 \zeta(3) g^4 }{576\pi^4 T}\,;
 \label{eq:dim6 matching}
 \\[0.2cm]
 \alpha_{81} &= \frac{31\zeta(5) g^8}{2048 \pi^6 T}\,,&
 \alpha_{82} &= -\frac{31\zeta(5) g^4}{10240\pi^6 T^3}\,,&
 \beta_{81} &= -\frac{31 \zeta (5) g^2 }{10240 \pi ^6 T^4}\,,
 \\[0.2cm]
 \beta_{82} &= \frac{217 \zeta(5) g^4}{20480 \pi ^6 T^3}\,,&
 \beta_{83} &= \frac{279 \zeta(5) g^4}{20480 \pi ^6 T^3}\,,&
 \beta_{84} &= -\frac{217 \zeta(5) g^6 }{5120 \pi ^6 T^2}\,.
  \label{eq:dim8 matching}
\end{align}
The terms in the first line were first computed in Ref.~\cite{Gould:2023jbz}, with which we find full agreement. 

Some of the operators above are physically equivalent, meaning that they can be related to one another via field redefinitions~\cite{Criado:2018sdb}. In accordance with our power counting, these redefinitions can be taken to be perturbative. In practice, operators of the form $f(\varphi)\partial^2\varphi$, that we label with $\beta$, can be removed from the Lagrangian through the transformation $\varphi\to\varphi+f(\varphi)$, at the price of shifting the coefficients $\alpha$.

Proceeding this way, with the help of \texttt{Matchete}~\cite{Fuentes-Martin:2022jrf}, we find that the shifts in the coefficients after all $f(\varphi)\partial^2\varphi$ operators have been eliminated are:\footnote{We have cross-checked these results with on-shell amplitude techniques in which the Lagrangians before and after field redefinitions are required to give exactly the same 4D S-matrix; see section 3.6 of Ref.~\cite{Aebischer:2023nnv} as well as Ref.~\cite{Chala:2024xyz}.}
\begin{align}
    \label{eq:initial}
    m_3^2&\to m_3^2 + 2\beta_{61}m_3^4 + (8\beta_{61}^2+2\beta_{81})m_3^6\,,
    \\[10pt]
    \kappa_3&\to \kappa_3 + \kappa_3 \left[6\beta_{61} m_3^2 + (42\beta_{61}^2+9\beta_{81})m_3^4\right]\,,
    \\[10pt]
    \lambda_3&\to\lambda_3 + 9\beta_{61}\kappa_3^2 + \left[\beta_{62}+30 (5\beta_{61}^2+\beta_{81})\kappa_3^2+8\beta_{61}\lambda_3\right]m_3^2
    \nonumber\\
    &\phantom{\to}+\left[10\beta_{61}\beta_{62}+\beta_{82}
    +\beta_{83}+4 (16\beta_{61}^2+3\beta_{81})\lambda_3\right]m_3^4\,,
    \\[10pt]
    \alpha_{61}&\to\alpha_{61}+4\beta_{62}\lambda_3
    +16\beta_{61}\lambda_3^2+\frac{3}{2}\kappa_3^2\left[78\beta_{61}\beta_{62}+9\beta_{82}+6\beta_{83}+8\lambda_3 (68\beta_{61}^2+13\beta_{81})\right]\nonumber\\
    &\phantom{\to}+\frac{1}{5}m_3^2 \left[60\alpha_{61}\beta_{61}+22\beta_{62}^2+5\beta_{84}+8 \lambda_3 (64\beta_{61}\beta_{62}+7\beta_{82}+5\beta_{83})
    \right.
    \nonumber\\
    &\phantom{\to+\frac{1}{5}m_3^2\big[}\left.+16\lambda_3^2 (108\beta_{61}^2+19\beta_{81})\right]\,,
    \\[10pt]
    \alpha_{81}&\to\alpha_{81}+6\alpha_{61}\beta_{62} + \frac{4}{5}\lambda_3(60 \alpha_{61}\beta_{61} + 27\beta_{62}^2 + 5 \beta_{84})
    \nonumber\\
    &\phantom{\to}+ \frac{16}{5} \lambda_3^2 (78 \beta_{61}\beta_{62} + 9 \beta_{82} + 5 \beta_{83}) + \frac{192}{5} \lambda_3^3 (16\beta_{61}^2+3\beta_{81})\,,
    \\[10pt]
    \alpha_{82} &\to\alpha_{82}\,.
\end{align}
On top of these, the $\mathbb{Z}_2$-breaking terms  $\alpha_5\varphi^5$ and $\alpha_7\varphi^7$ appear, with:
\begin{align}
\alpha_5 &= 3\kappa_3 \left[\beta_{62}+12 (5\beta_{61}^2+\beta_{81})\kappa_3^2+8\beta_{61}\lambda_3\right]\nonumber\\
&\phantom{=}+\frac{3}{2}\kappa_3 (46\beta_{61}\beta_{62}+5\beta_{82}+4\beta_{83}+304\beta_{61}^2\lambda_3+56\beta_{81}\lambda_3)m_3^2\,,
\\[10pt]
\alpha_7 &= \frac{3}{10}\kappa_3 \left[120\alpha_{61}\beta_{61}+49\beta_{62}^2+10\beta_{84}+4\lambda_3 (286\beta_{61}\beta_{62}+33\beta_{82}+20\beta_{83})\right. 
\nonumber\\
&\phantom{=}\left.+256\lambda_3^2(16\beta_{61}^2+3\beta_{81})\right]\,.\label{eq:final}
\end{align}
(Further trivial-to-compute corrections arise first upon normalising canonically the kinetic term, $\varphi\to \varphi/\sqrt{K_3}$, perturbatively.) This basis is particularly useful because it minimises the number of operators with derivatives, which entail most numerical complications when computing the classical bounce solutions.

\begin{figure}[t]
\includegraphics[width=\linewidth]{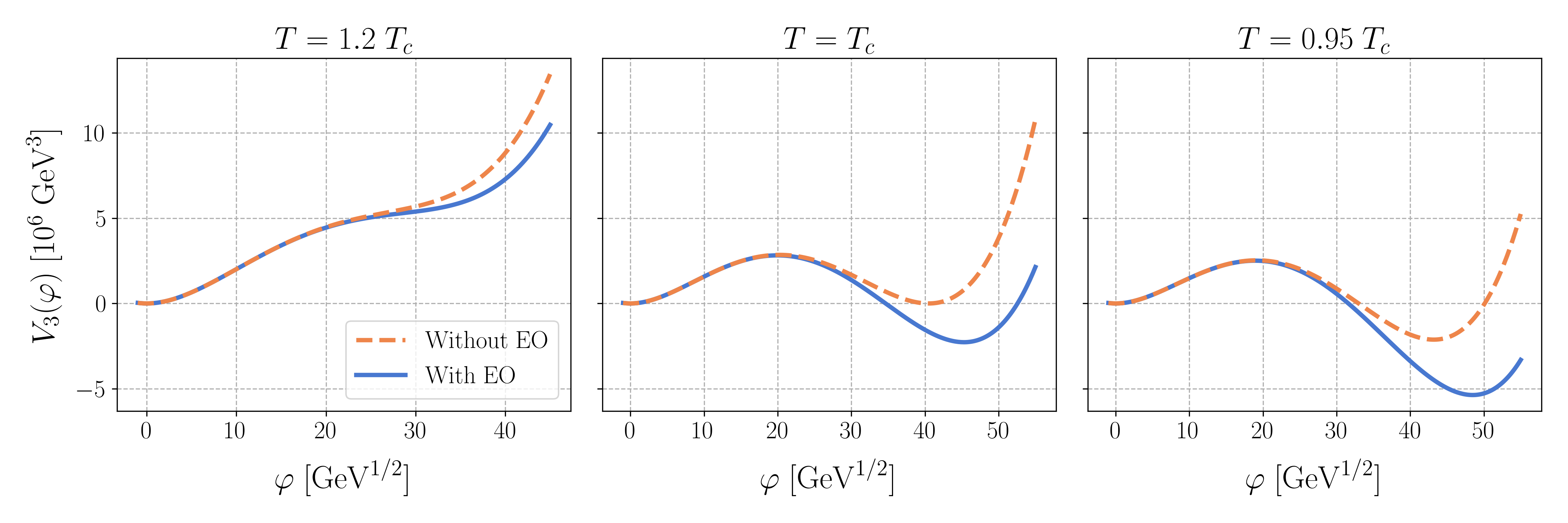}
 \caption{\it Effective potential below (left), at (middle) and above (right) $T_c=382.84 ~\mathrm{GeV}$, as obtained for the potential without effective operators. The solid and dashed lines represent the situation with and without effective operators (EO), respectively; see the text for details.}\label{fig:potentials}
\end{figure}

In Fig.~\ref{fig:potentials}, we plot the effective potential (that is, the derivative-independent part of $\mathcal{L}_3$) at $T>T_c$ (left), $T=T_c$ (middle) and $T<T_c$ (right). Here, $T_c$ stands for the critical temperature, namely that at which the effective potential exhibits two degenerated minima, $\varphi_F$ and $\varphi_T$, computed in the absence of effective interactions. We have chosen a model with $\left(m^2, \kappa, \lambda\right) = \left(31\,643.5~\mathrm{GeV^2}, -71.1~\mathrm{GeV}, 0.045 \right)$, which is representative of the trend we observe in a variety of scanned models \mc{and that we discuss in section~\ref{sec:results} (namely, that including higher-order-operator corrections modifies significantly the range of values of $g$ allowing for a PT, as well as the estimation of GW-related magnitudes in said range).}

\section{Phase-transition parameters}
\label{sec:parameters}
PTs proceed through the nucleation of bubbles of vacuum energy~\cite{Langer:1969bc,Langer:1974cpa}. These grow and eventually collide, producing a stochastic background of GWs~\cite{Caprini:2019egz}. The main physical parameters describing this dynamics are the nucleation temperature ($T_*$), the latent heat ($\alpha$), the inverse duration time of the PT ($\beta/H_*$) and the terminal bubble wall velocity ($v_w$).

Before entering into the definition of these quantities, let us mention that they are all determined (at the classical level in the EFT) by the value of the 3D effective action at classical solutions of the equations of motion (EOM), $\varphi_c$. The latter can either be constant, which describes the extrema of the effective potential, or a bounce; namely a non-homogeneous but spherically-symmetric field configuration that interpolates between $\varphi_F$ and $\varphi_T$~\cite{Coleman:1977py,Linde:1980tt}.
We will study here the effects of higher-dimensional EFT operators on the bounce solution and the corresponding action. This amounts to the inclusion of higher orders in the expansion in powers of $P/T \sim m/T \sim g$. We will neglect additional corrections coming from loops of the Matsubara zero modes.

At high temperature~\cite{Ghiglieri:2020dpq,Bodeker:2022ihg}, for an action of the form
\begin{equation}\label{eq:trivialaction}
    S_3[\varphi] = 4\pi \int_0^\infty dr \, r^2 \left[\frac{1}{2}\dot{\varphi}^2+V_3(\varphi)-V_3(\varphi_F)\right]\,,
\end{equation}
where the dot stands for derivative with respect to $r$, the bounce is a solution of the corresponding Euler-Lagrange equation:
\begin{equation}
    \ddot{\varphi}+\frac{2}{r}\dot{\varphi} = V_3'(\varphi)\,,
    \label{eq:bounce-eq-wo-op}
\end{equation}
with boundary conditions $\dot{\varphi}(0)=0$ and $\lim_{r\to\infty}\varphi(r)=\varphi_F$.
Without loss of generality, we assume from now on that $\varphi_F = 0$.

The existence of bounce solutions in theories with an arbitrary potential with degenerate minima was proven by Coleman in 1977~\cite{Coleman:1977py}. In the presence of other derivative interactions, however, it has not been proven whether such a solution exists. As a matter of fact, to the best of our knowledge, none of the dedicated tools for computing the bounce~\cite{Wainwright:2011kj,Masoumi:2016wot,Athron:2019nbd,Sato:2019wpo,Guada:2020xnz} accepts derivative interactions other than the usual kinetic term.

Even if the bounce could be obtained directly by an extension of the conventional methods, in doing so one would lose track of the EFT expansion.
While this is not a problem in principle, it would make it impossible to test for the breaking point of perturbativity by comparing different orders, as we do below.
Additionally, we would like to check that the bounce actions before and after the redefinitions performed in Section~\ref{sec:theory} are equal.
If each order in perturbation theory can be clearly identified, this is easy to do, as they should be exactly equal.
Otherwise, one can only access the full actions, which will be only approximately equal (up to effects of dimension-10 operators that we have neglected).
This resembles very much the gauge dependence of the effective potential at its extrema if not computed consistently in perturbation theory~\cite{Patel:2011th,Ekstedt:2018ftj,Ekstedt:2020abj}

To compute the physical $S_3[\varphi_c]$ in a way consistent with the perturbative expansion used for the matching, we rely on expanding both the classical bounce and the action in powers of $\epsilon$, a formal parameter that keeps track of the perturbative order: 
\begin{align}
    \varphi_c = \varphi_c^{(0)} + \epsilon\varphi_c^{(1)}+\epsilon^2\varphi_c^{(2)}+\cdots\,,\quad
    S_3 = S_3^{(0)}+\epsilon S_3^{(1)} + \epsilon^2 S_3^{(2)}+\cdots
    \label{eq:bounce-action-perturbative}
\end{align}
Here, $\epsilon$ is defined such that the terms in the first row of Eq.~\eqref{eq:3deft} are $\epsilon$-independent, while those in the second and third rows come with $\epsilon$ and $\epsilon^2$, respectively. At the end of our calculations, we take $\epsilon = 1$.

Requiring the bounce to be an extremal of $S_3$, namely $\frac{\delta}{\delta\varphi} S_3|_{\varphi_c}=0$, we obtain:
\begin{align}\label{eq:physicalbounce}
    S_3[\varphi_c] = S_3^{(0)}[\varphi_c^{(0)}]+\epsilon S_3^{(1)}[\varphi_c^0] + \epsilon^2\left\lbrace S_3^{(2)}[\varphi_c^{(0)}]+2\pi\int_0^\infty dr \, r^2 \varphi_c^{(1)} \frac{\delta S_3^{(1)}}{\delta\varphi}\bigg|_{\varphi_c^{(0)}}\right\rbrace + \mathcal{O}(\epsilon^3)
\end{align}
where $\varphi_c^{(0)}$ is the bounce of the zeroth-order action, and with $\varphi_c^{(1)}$ satisfying the following differential equation:
\begin{align}
       \ddot{{\varphi}}_c^{(1)}+\frac{2}{r}\dot{\varphi}_c^{(1)}-V_3^{(0)''}(\varphi_c^{(0)}) ~\varphi_c^{(1)}-\frac{1}{4\pi r^2}\frac{\delta S_3^{(1)}}{\delta\varphi}\bigg|_{\varphi_c^{(0)}}&=0\
    \label{eq:diff equation for phi1 model}
\end{align}
with boundary conditions
\begin{equation}
 \dot{\varphi}_c^{(1)}(0)=\lim_{r\to\infty}\varphi_c^{(1)}(r)=0.
\end{equation}
The mathematical details of this derivation, to arbitrary order in $\epsilon$, are discussed in Appendix~\ref{app:bounce}. Similar ideas have been used to compute gauge-invariant nucleation rates~\cite{Lofgren:2021ogg,Hirvonen:2021zej} as well as for obtaining the functional derivative of the effective action at zero temperature~\cite{Baacke:1995hw,Metaxas:1995ab}.

\begin{figure}[t]
\includegraphics[width=0.49\linewidth]{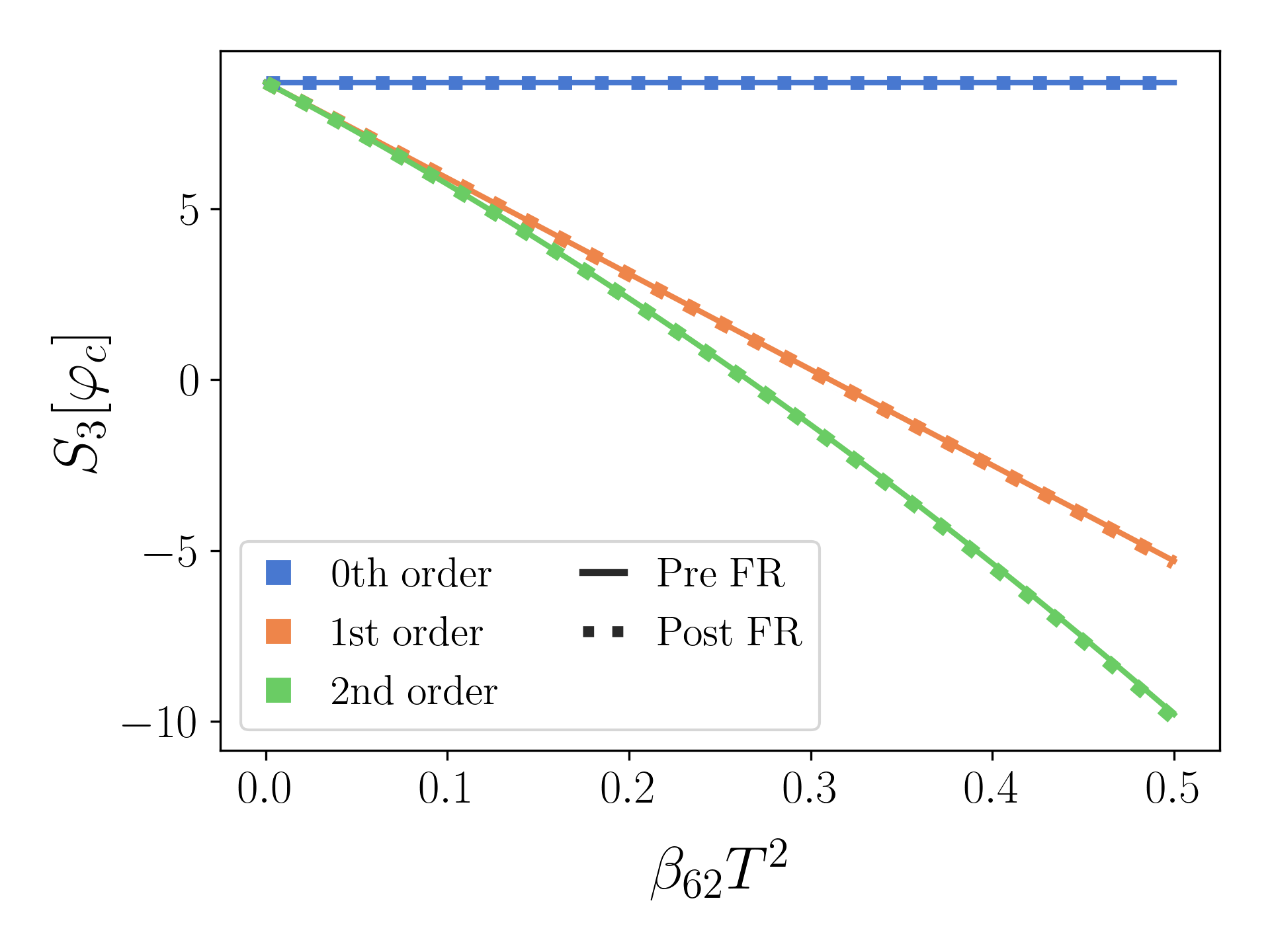}
\includegraphics[width=0.49\linewidth]{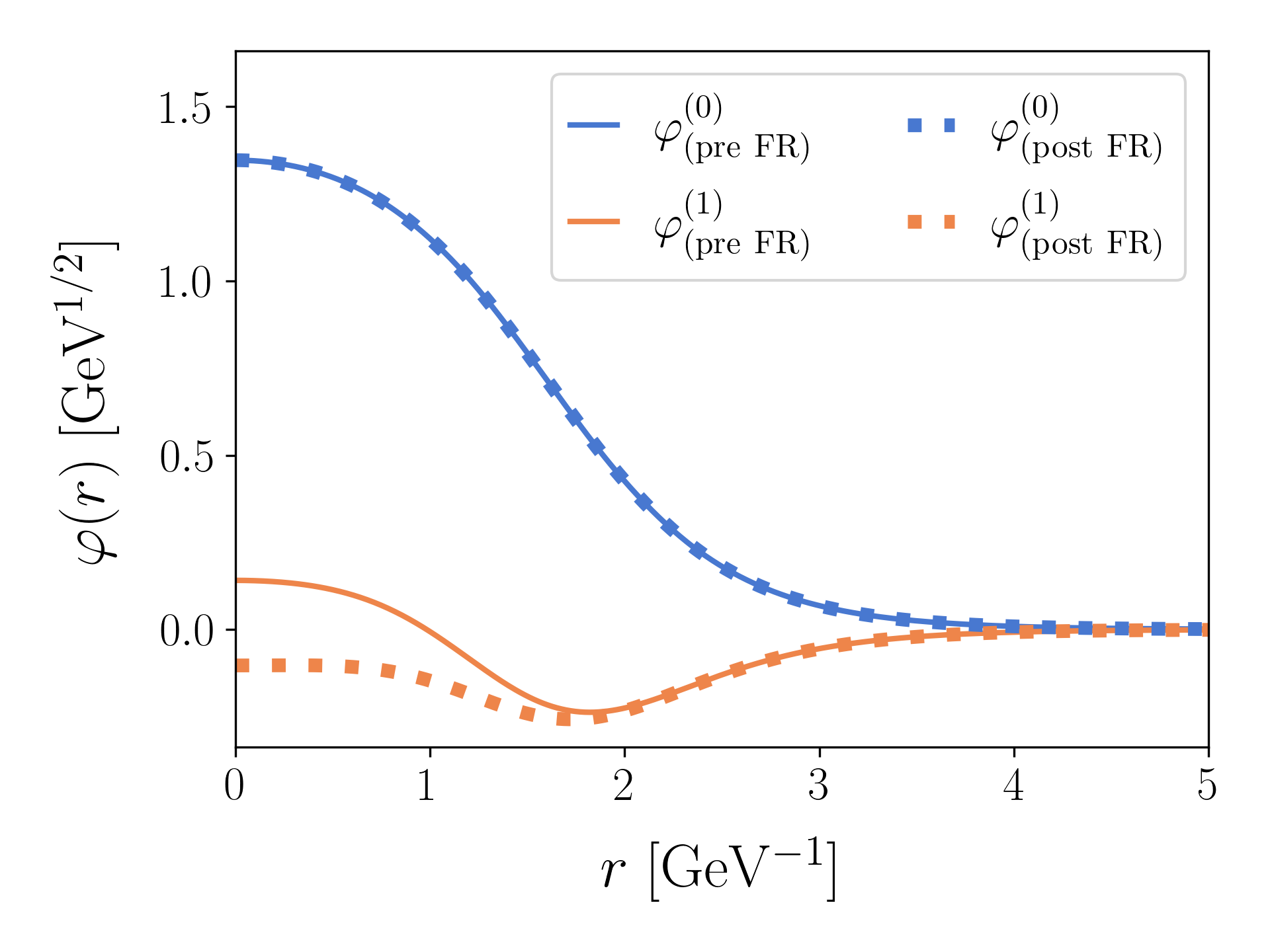}
\caption{\it Results before and after field redefinitions. Left: Value of $S_3[\varphi_c]$ computed up to different orders in the EFT power counting. Right: Leading bounce and first correction; see the text for details.}\label{fig:bounces}
\end{figure}

In order to further emphasise that $S_3[\varphi_c]$ computed this way is invariant under field redefinitions, in Fig.~\ref{fig:bounces}, we show explicitly the value of this quantity in a scenario with only non-vanishing Wilson coefficient $\beta_{62}$ before and after the field redefinitions defined by Eqs.~\eqref{eq:initial}--\eqref{eq:final}. We also plot $\varphi_c^{(0)}$ and $\varphi_c^{(1)}$. While the redefinition changes $\varphi_c^{(1)}$, physical observables are invariant under it, since they depend on $\varphi_c^{(1)}$ through the action $S_3[\varphi_c]$, which does not change. We have chosen $K_3=1$, $m_3^2=3.20~\mathrm{GeV}^2$, $\kappa_3=-3.40~\mathrm{GeV^{3/2}}$ and $\lambda_3 =1.34~\mathrm{GeV}$. Notice that, since there is only one operator in this case, we have taken $\epsilon = \beta_{62} T^2$ as our perturbative parameter.

With this information, assuming that the PT occurs in a radiation dominated epoch, a precise definition of $T_*$, $\alpha$ and $\beta/H_*$ follows:\footnote{We remind the reader that these quantities are defined using the 3D Euclidean action, potential and field, and thus have been adapted from the usual definitions in terms of their 4D counterparts so as to reproduce the appropriate energy dimensions.}

\begin{itemize}
 \item $T_*$: It is defined as the $T$ at which the probability $\mathcal{P}\sim (M_{\rm Pl} / T)^4 \, e^{-S_3(T)}$ for a single bubble to nucleate within a Hubble horizon volume is $\sim 1$. Numerically,
 this occurs when~\cite{Quiros:1999jp}
 \begin{equation}
  S_3[\varphi_c]\sim 100-4\log{\frac{T_*}{100\,\text{GeV}}}\,.
  \label{eq:Tnuc}
 \end{equation}
 \item $\alpha$: It is defined as the ratio of the trace anomaly difference of the energy momentum tensor between the phases to the energy density of the radiation bath. Taking the effective number of degrees of freedom in the plasma as determined in the SM, we have~\cite{Athron:2023xlk}: 
 \begin{equation}
  \alpha = \frac{\left.\Delta \left( V_3(\varphi) - \frac{T}{4} \frac{\mathrm{d}}{\mathrm{d}T} V_3(\varphi) \right)\right|_{T_*}}{\rho_r(T_*)} \thickapprox -0.03 \dfrac{\left.\Delta \left( V_3(\varphi) - \frac{T}{4} \frac{\mathrm{d}}{\mathrm{d}T} V_3(\varphi) \right)\right|_{T_*}}{T_*^3}\,,
 \end{equation}
 where  $\rho_r = g(T) \pi^2 T^4/30$ is the radiation energy density of the plasma.
 \item $\beta/H_*$: It is defined as a characteristic timescale of the PT, assuming an exponentially growing transition rate as the temperature decreases (or equivalently, after linearising the bounce action with respect to the temperature) ~\cite{Caprini:2019egz}:
 \begin{equation}
  \frac{\beta}{H_*} = T_* \frac{\text{d}S_3[\varphi_c]}{\text{d}T}\bigg|_{T_*}\,.
 \end{equation}
\end{itemize}
One last parameter, the bubble wall velocity $v_w$, enters also into the determination of the GWs produced during the PT. For its computation, we use the approximate formula \cite{Lewicki:2021pgr}
\begin{equation}
    v_w =
    \begin{cases}
        \sqrt{\frac{\Delta V}{\alpha \rho_r}} & \text{for} \quad \sqrt{\frac{\Delta V}{\alpha \rho_r}} < v_J(\alpha) \\
        1 & \text{for} \quad \sqrt{\frac{\Delta V}{\alpha \rho_r}} \geq v_J(\alpha)
    \end{cases}\,,
    \label{eq:vw}
\end{equation}
where $\Delta V = V_3(\varphi_T)$ is the difference in the potential between the phases, 
and $v_J$ is the Jouguet velocity, defined as
\begin{equation}
    v_J = \frac{1}{\sqrt{3}} \frac{1 + \sqrt{3 \alpha^2 + 2 \alpha}}{1 + \alpha}\,.
\end{equation}

We note in passing that, just like for $S_3[\varphi_c]$, $V_3(\varphi_T)$ must be computed perturbatively. That is:
\begin{align}
V_3(\varphi_T) = V_3^{(0)}(\varphi_T^{(0)}) + 
\epsilon V_3^{(1)} (\varphi_T^{(0)}) 
+ \epsilon^2 \left\{ V_3^{(2)}(\varphi_T^{(0)}) - \frac{1}{2} \frac{\left(V_3^{(1) '}(\varphi_T^{(0)})\right)^2}{V_3^{(0) ''}(\varphi_T^{(0)})}\right\}
+ \mathcal{O}(\epsilon^3)\,.
\end{align}
Since $S_3[\varphi_c]$ is computed perturbatively, the nucleation temperature $T_*$ as defined in Eq.~\eqref{eq:Tnuc}, is also field-redefinition invariant. This, together with a correct perturbative estimation of $V_3(\varphi_T)$, guarantees that the strength parameter $\alpha$ is also field-redefinition invariant, even if this parameter is not precisely expanded perturbatively as a whole. In any case, errors in the estimation of $\alpha$ are smaller than $\mathcal{O}(\epsilon^2)$ in our formal power counting. 

\section{Results}
\label{sec:results}
\begin{figure}[t]
 \includegraphics[width=0.49\linewidth]{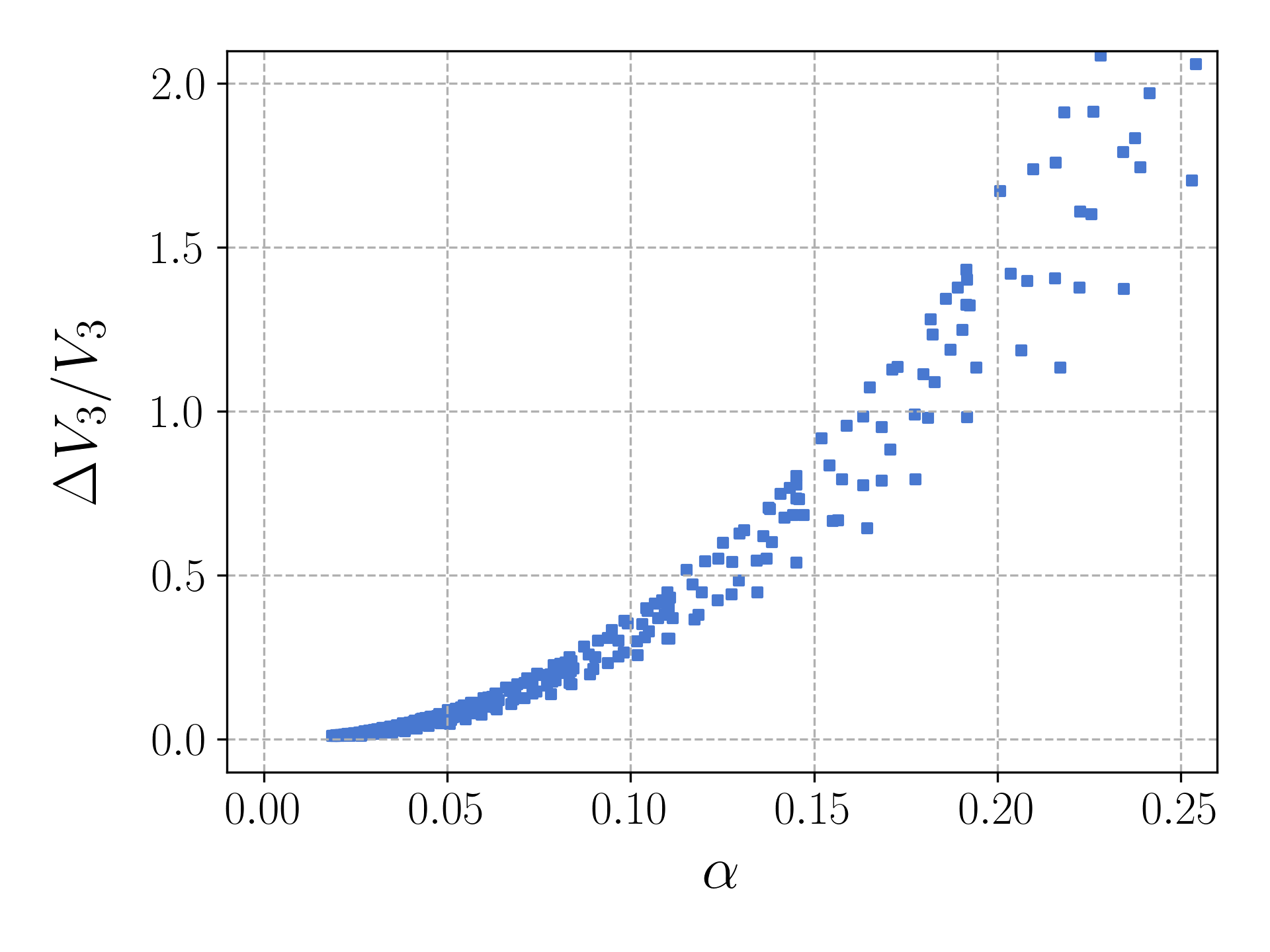}
 \includegraphics[width=0.49\linewidth]{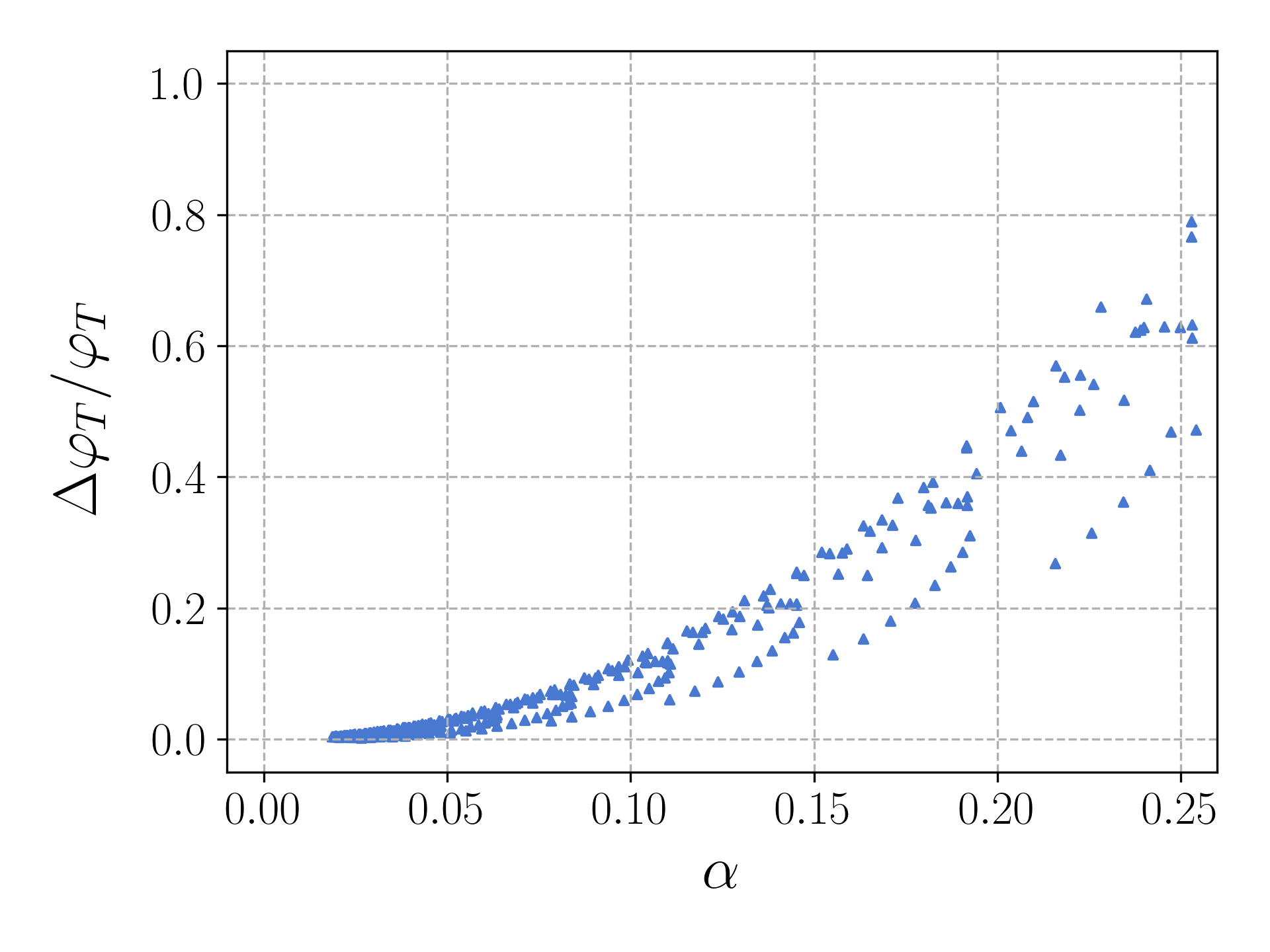}
 \caption{\it Scatter plots with the minimum corrections for $\Delta V_3/V_3 \equiv V_3^{(1)}(\varphi_T^{(0)})/V_3^{(0)}(\varphi_T^{(0)})$ and $\Delta \varphi_T \equiv \varphi_T^{(1)}/\varphi_T^{(0)}$ for several models; see the text for details.}\label{fig:proxy}
\end{figure}
%
%

In order to get a first indication of the impact of higher-dimensional operators on the aforementioned parameters, we proceed as follows. We take the results of Ref.~\cite{Chala:2019rfk}, which provides a thorough scan of a simplified 3D model without effective interactions, given by
\begin{equation}
    \mathcal{L}_3 = \frac{1}{2} (\partial \varphi)^2 + \frac{1}{2} \left( m^2 + \frac{g^2 T^2}{6} \right) \varphi^2 + \kappa \sqrt{T} \varphi^3 + \lambda T \varphi^4\,.
\end{equation}
This scan includes values of $\alpha$, $m^2_3$, $\kappa_3$ and $\lambda_3$ at $T_*$
Given $g$, then $m^2$, $\kappa$ and $\lambda$ can be obtained straightforwardly.
For each point with $\alpha>0.1$, $m/(\pi T_*)<1$ and $\kappa/(\pi T_*)<1$, namely a strong PT within the regime of validity of the EFT, we compute the minimum value of $g$ for which $\beta/H_*>0$. Negative values of this magnitude point out to a breaking of the assumption that the transition rate grows exponentially with $T$. These scenarios and their relation to supercooled PTs have recently been discussed in the literature \cite{Athron:2023}. In this work, we shall only consider PTs where $\beta/H_* > 0$.

\begin{figure}[t]
\centering
\includegraphics[width=0.49\linewidth]{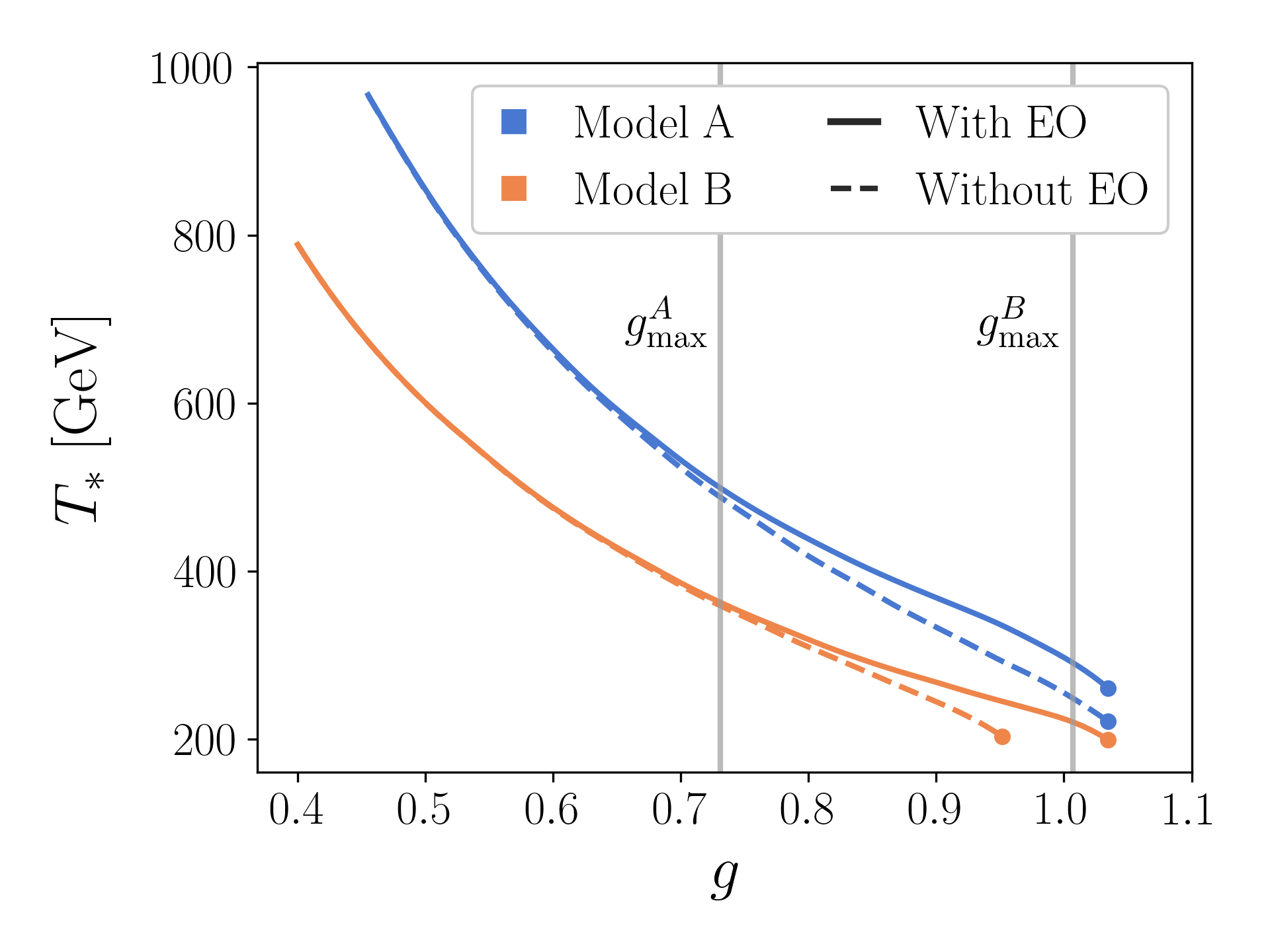}
\includegraphics[width=0.49\linewidth]{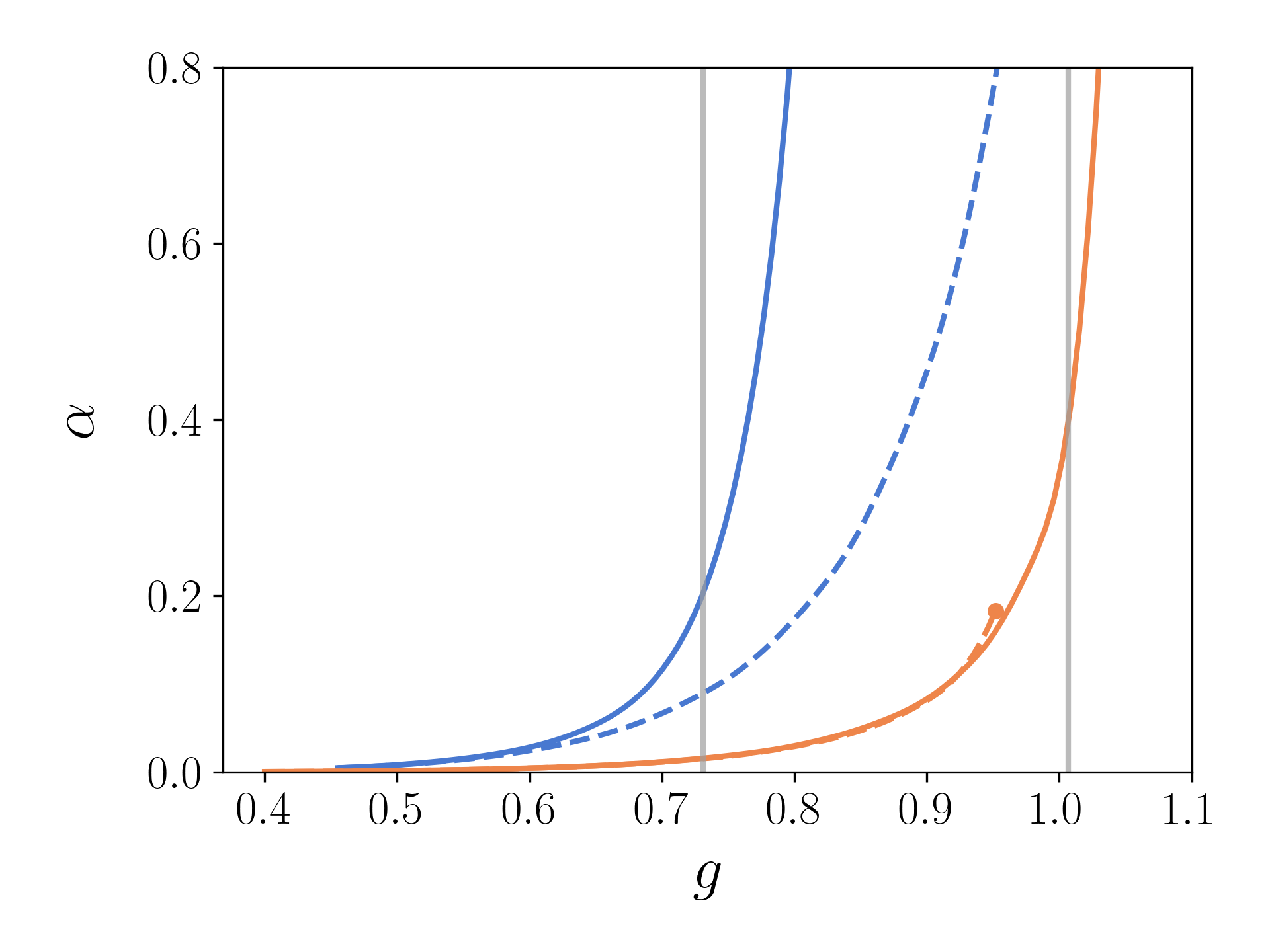}

\includegraphics[width=0.49\linewidth]{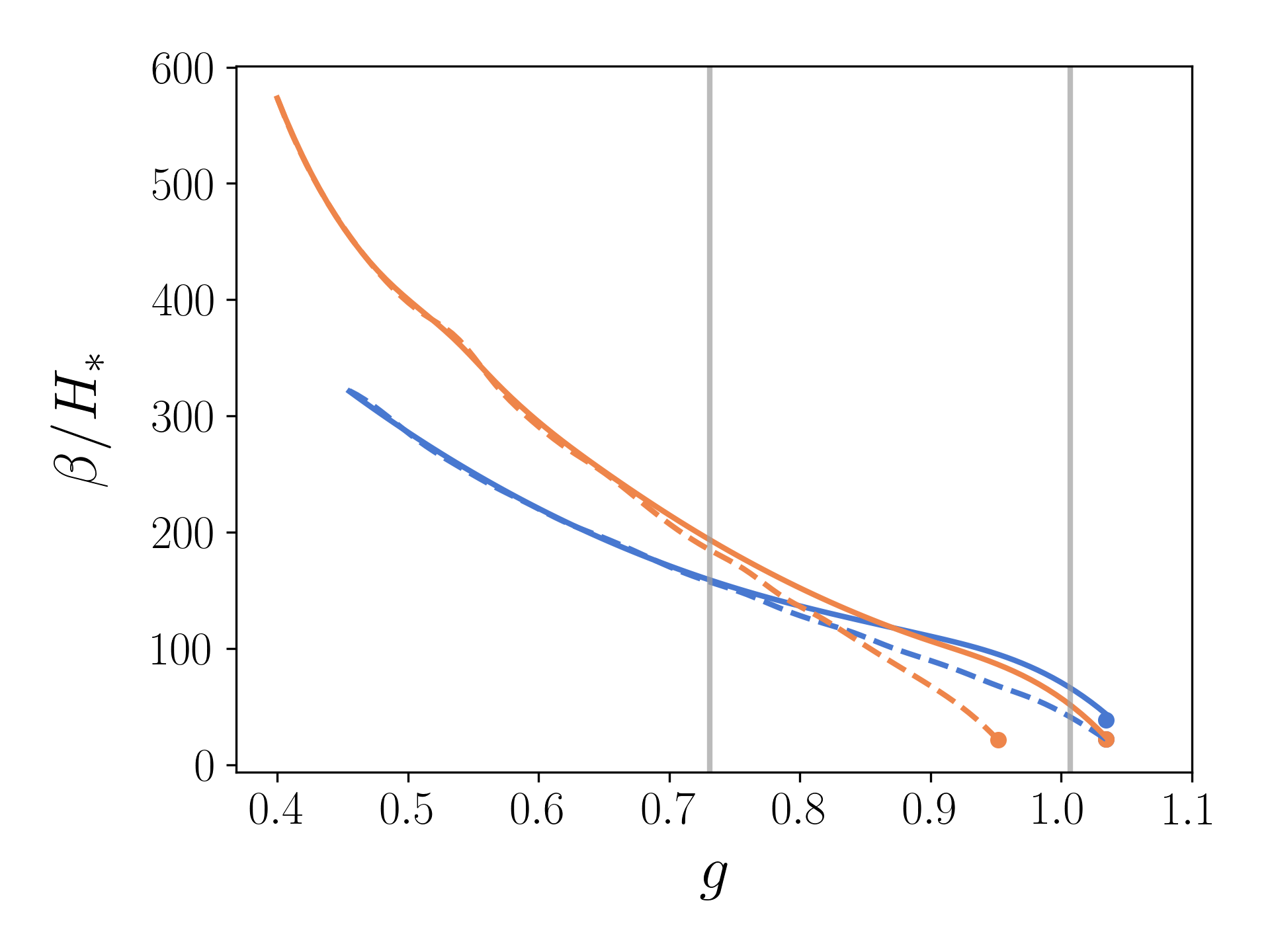}
\includegraphics[width=0.49\linewidth]{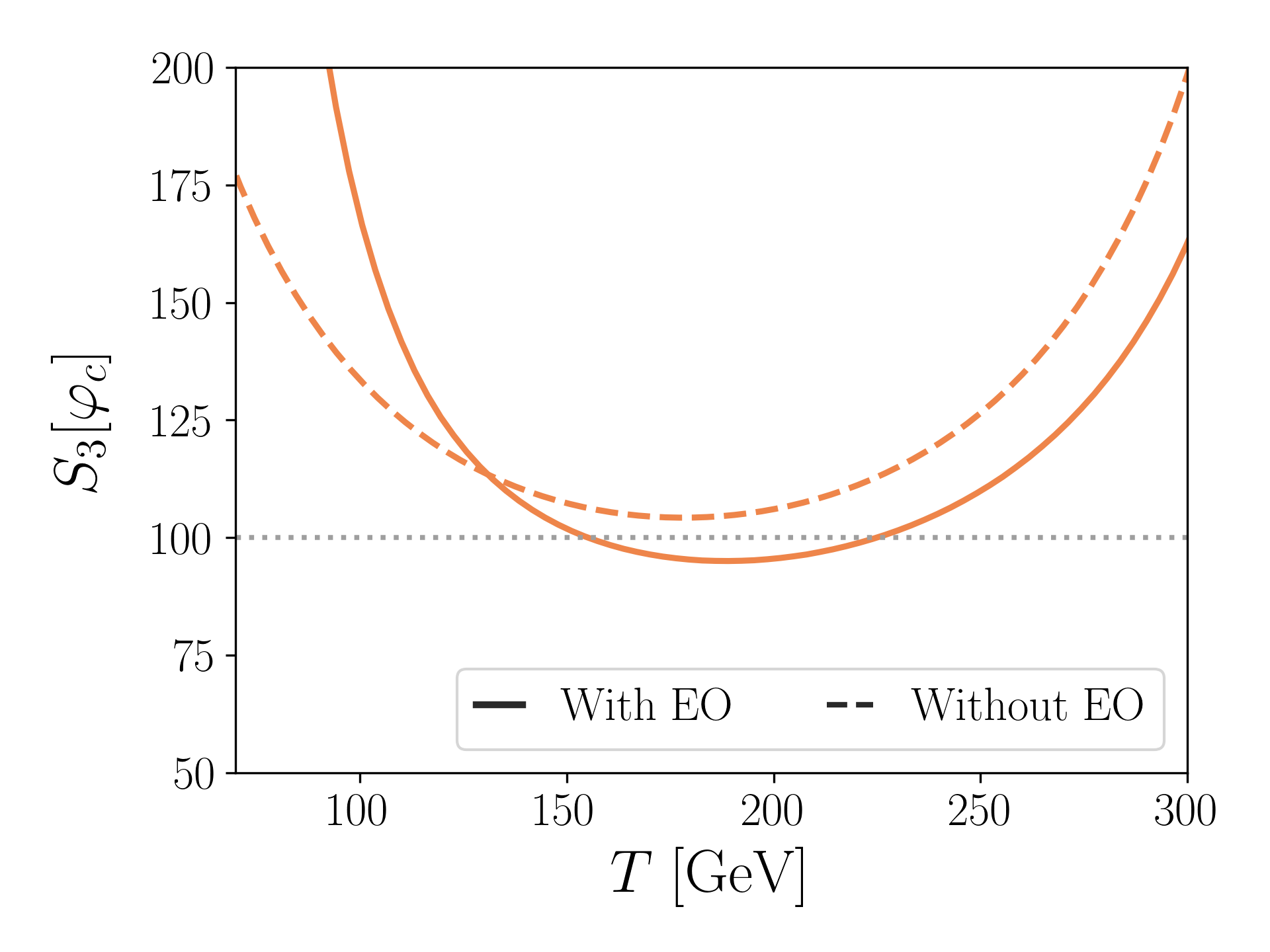}
 \caption{\it Several plots comparing PT parameters with and without effective operators (EO): curves of $T_*$ (top left), $\alpha$ (top right) and $\beta/H_*$ (bottom left) as a function of $g$. We use two different models, with parameters $(m^2, \kappa, \lambda)_A = (20\,000 ~\mathrm{GeV^2}, -40 ~\mathrm{GeV}, 0.01)$ and $(m^2, \kappa, \lambda)_B = (31\,643.5 ~\mathrm{GeV^2}, -71.1 ~\mathrm{GeV}, 0.045)$. In the bottom right, we plot the curve of the perturbative bounce action as a function of $T$ with and without effective operators for model B and $g=1.0$, where the PT does not occur if these interactions are not considered. The vertical lines represent $g_{\rm max}$, defined as the value of $g$ at which $V_3^{(2)}(\varphi^{(0)}_T)/V_3^{(1)}(\varphi^{(0)}_T)\sim 0.5$, implying a (safe) limit on the validity of the EFT. 
 }
 \label{fig:tnalphabeta}
\end{figure}

Next, we include an effective interaction term $\varphi^6$ to the potential, with its Wilson coefficient given by the matching Eq.~\eqref{eq:dim6 matching}. For the minimum $g$, we compute the (minimum) value of $V_3^{(1)}(\varphi_T^{(0)})/V_3^{(0)}(\varphi_T^{(0)})$ as well as the shift in $\varphi_T$ after including this effective operator. These quantities (both related to the value of the strength parameter $\alpha$) give an indication of how large effective-operator corrections can be. They are shown in Fig.~\ref{fig:proxy}. It is evident that large values of $\alpha$ correlate with large corrections in both cases. 
Returning to the complete model in Eq. \eqref{eq:3deft}, we use the same 4D Lagrangian parameters extracted from the scan and we compute $T_*$, $\alpha$, $\beta/H_*$ and $v_w$ as functions of $g$ taking correctly into account the effective interactions. We show the first three magnitudes in two representative cases in Fig.~\ref{fig:tnalphabeta}. We do not plot $v_w$ as its value is straightforwardly determined by $\alpha$, as shown in Eq. \eqref{eq:vw}. \mc{Unless stated otherwise, all results are computed at the matching scale $\Lambda = \pi T e^{-\gamma}$, where $\gamma$ is the Euler-Mascheroni constant.}

The plots represent a common trend we observe: in most cases, the PT takes place for a wider range of values of $g$ (generally allowing for larger values of $\alpha$) when effective operators are included. To inspect this finding in more detail, we also show the evolution of $S_3[\varphi_c]$ as a function of $T$ for a value of $g$ in model B, where including effective operators allows for PTs for a wider range of $g$.\footnote{
This trend can be roughly understood on the basis of Fig.~\ref{fig:potentials}: the effective terms tend to lower the true vacuum with respect to the false one, making the PT possible. }

\begin{figure}[t]
    \centering
    \includegraphics[width=0.49\linewidth]{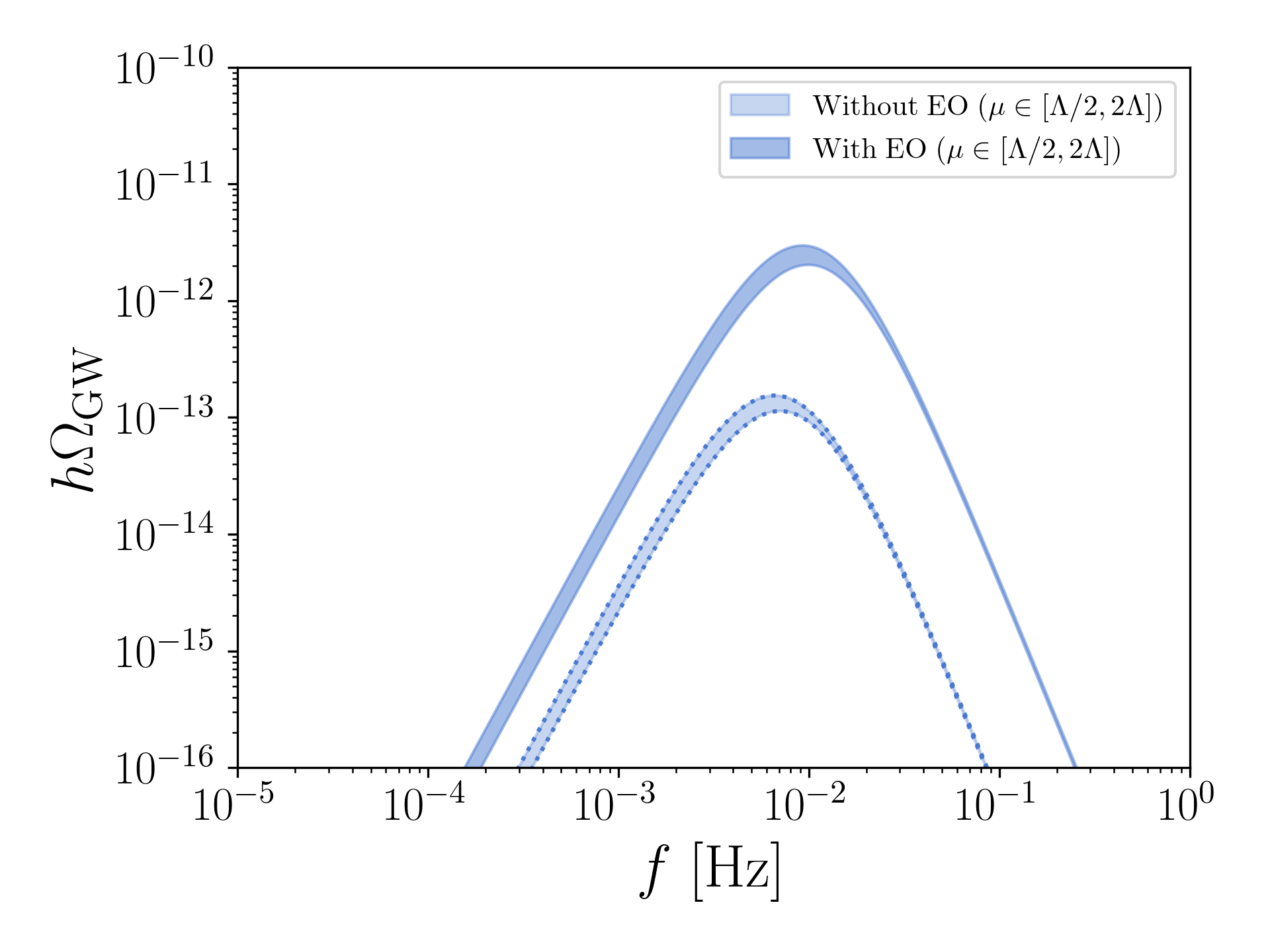}
    \includegraphics[width=0.49\linewidth]{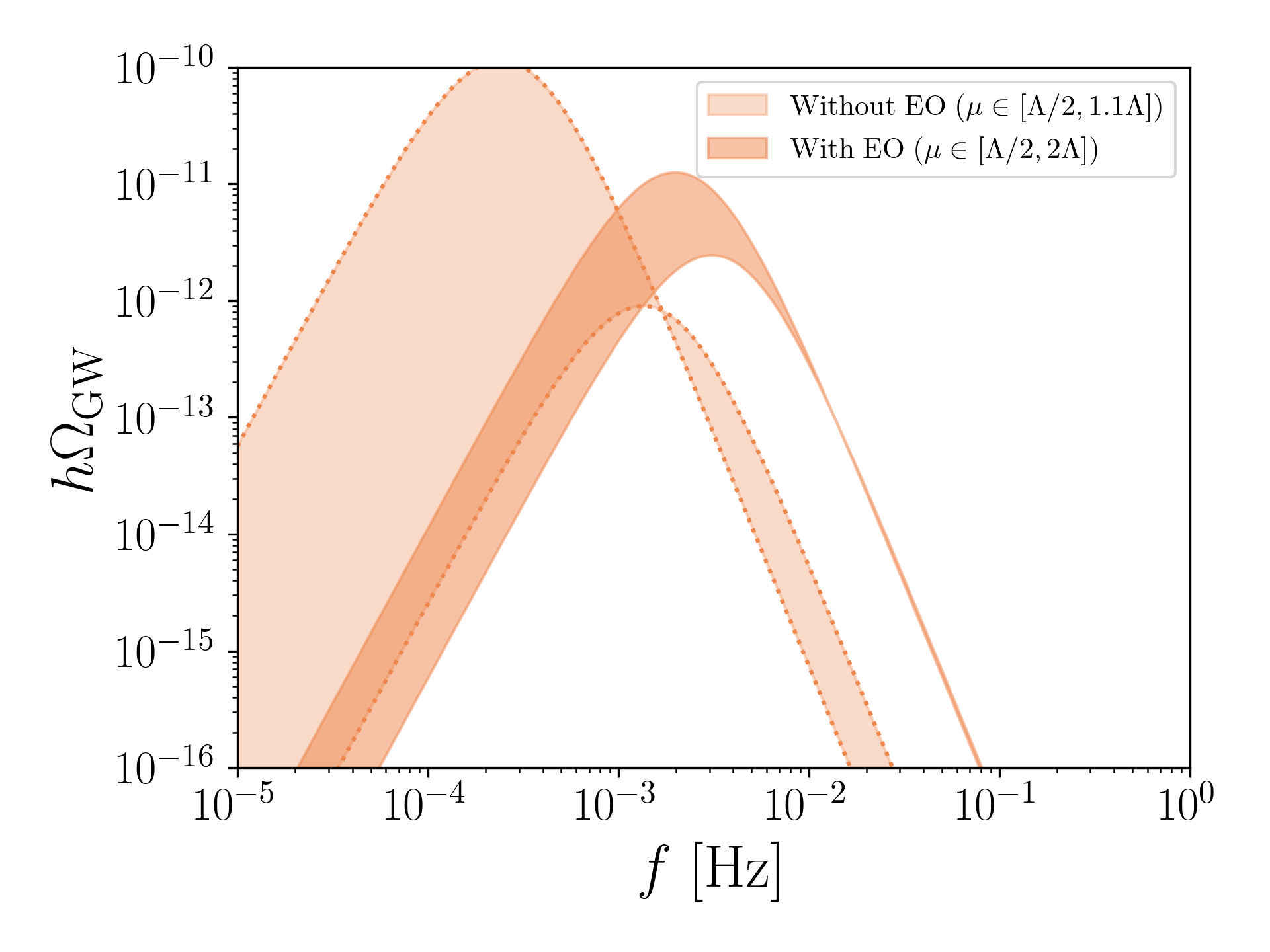}
    \caption{\it Left: Sound wave gravitational wave power spectra for model A (see caption in Fig. \ref{fig:tnalphabeta}) at $g_\mathrm{max}^A \sim 0.73$, for which we obtained $v_w \sim 0.74$ and $v_w = 1$ including and not including effective operators (EO), respectively. Right: Same, but for model B at $g\sim 0.95$, close to the maximum $g$ such that a PT occurs without effective operators. In this case $v_w \sim 0.74$ and $v_w = 1$, with and without EO, respectively; see the text for details.
    }    
    \label{fig:gws}
\end{figure}

It should be now clear that the effective interactions tend to bend the curve of bounce action versus $T$, so that it now reaches the value of $\sim 100$ in a larger range of $g$. As we anticipated from the preliminary test in Fig. \ref{fig:proxy}, it is also apparent that, when $\alpha$ is relatively large, the predictions for the PT parameters with and without effective operators can be drastically different. This, in turn, has an enormous impact on the GW power spectrum resulting from these PTs. To show this, in Fig.~\ref{fig:gws} we provide plots with the predicted GWs in two different parameter space points. To this aim, we use the equations for the dominant (sound wave) contribution for a given frequency $f$, as taken from the technical note for \texttt{PTPlot}~\cite{PTPlot}. 
%

In the plots we observe that the peak amplitude of the power spectrum \mc{can change by more than} one order of magnitude in both models, while the peak frequency is not significantly modified for model A but \mc{can shift}  by a factor of $10$ for model B. The large differences we observe are due to the notable disagreement in the estimation of the four relevant PT parameters when including effective operators: $T_*$, $\alpha$, $\beta/H_*$ and $v_w$. For model A we find that $|\Delta T_*/T_*| \thickapprox 0.02$, $|\Delta \alpha / \alpha| \thickapprox 0.56$, $|\Delta \beta / \beta| \thickapprox 0.01$ and $|\Delta v_w / v_w| = 0.26$, while for model B we get $|\Delta T_*/T_*| \thickapprox 0.17$, $|\Delta \alpha / \alpha| \thickapprox 0.15$, $|\Delta \beta / \beta| \thickapprox 0.75$ and $|\Delta v_w / v_w| = 0.26$.

\mc{The bands represent the uncertainty ensuing from running the matching scale $\mu$ in the range $[0.5\Lambda,2 \Lambda]$. (We also vary the 3D scale, but the impact is negligible.) Note however that, in model B, in the absence of higher-dimensional operators, there is no PT for values of $\mu\gtrsim 1.2\Lambda$, so in this case we only allow for values below this point.}


\section{Conclusions}
\label{sec:conclusions}
We have performed the dimensional reduction of a model with a real scalar and a massless fermion, including matching corrections to effective operators in the EFT with more than four scalars and/or more than two derivatives. The details of the matching computation can be found in Appendix~\ref{app:matching}. Subsequently, we have discussed how to compute the different PT parameters in the presence of the effective operators; see Appendix~\ref{app:bounce} for in-depth information.

On the basis of these results, we have analysed how the dynamics of strong PTs, here defined as having $\alpha\gtrsim 0.1$, change if effective interactions are not neglected. We have found that the dominant effect is that, in most such parameter space points, effective interactions allow for PTs in a wider range of values of the Yukawa coupling.
Equally interesting is that, there where PTs happen both in the presence and in the absence of EFT terms, the predictions for $T_*$, $\alpha$, $\beta/H_*$, $v_w$ and for the subsequent GWs \mc{can change} very significantly. 

Our work thus provides robust evidence of the importance of including EFT effects in the study of strong PTs. Since we have tried to isolate as much as possible the effects of effective interactions, we have not considered higher-loop corrections in the matching nor in the effective potential. Incorporating these effects in a comprehensive analysis of the parameter space of the model constitutes a possible future line of work. Other potentially interesting avenues comprise including quantum corrections (due to light fields) in the calculation of the effective action, in line with Refs.~\cite{Baacke:1993ne,Dunne:2005rt,Ai:2018guc,Ai:2020sru,Ekstedt:2023sqc,Ai:2023yce,Matteini:2024xvg}; or applying these methods to other well-motivated models and EFTs~\cite{Grojean:2004xa,Chala:2018ari,Camargo-Molina:2021zgz,Hashino:2022ghd,Alonso:2023jsi,Oikonomou:2024jms} for physics beyond the SM.

\section*{Acknowledgments}
We thank Jose Santiago for discussions about all aspects of the project. We are also grateful to Andreas Ekstedt for enlightening discussions and for important comments on the manuscript. We also thank Johan L\"ofgren for feedback and for spotting several typos.
This work has received funding from MICIU/AEI/10.13039/501100011033 and ERDF/EU (grants PID2022-139466NB-C22 and PID2021-128396NB-I00) as well as from the Junta de Andaluc\'ia grants FQM 101 and P21-00199. MC and JCC are further supported by the Ram\'on y Cajal program under grants  RYC2019-027155-I and RYC2021-030842-I; respectively.

\appendix

\section{Detailed computations of the matching}
\label{app:matching}

In this appendix we show the results for the matching conditions of the 1PI correlation functions between the UV (4D) and IR (3D) theories.

\subsection{Setup}

We compute off-shell $n$-point functions, with $1 \leq n \leq 8$, to one-loop order in the UV theory and match them onto the tree-level in the 3D counterpart. All diagrams with one-loop of fermions and an odd number of external legs vanish, as they involve taking the trace of an odd number of gamma matrices.

The Feynman rules for both theories are obtained in Minkowski space, with signature $\eta_{\mu\nu} = \mathrm{diag}~(+---)$ using \texttt{FeynRules} \cite{Alloul:2014}. In order to change to Euclidean space, with $g_{\mu\nu} = \delta_{\mu\nu}$, we proceed as follows: 
\begin{enumerate}
    \item Split off a factor of $i$ from every vertex.
    \item Change the scalar product to Euclidean space: $a^\mu b_\mu \to -a_E \cdot b_E$.
    \item Change propagators from Minkowski to Euclidean space: $\dfrac{i}{p^2 - m^2} \to \dfrac{1}{P^2 + m^2}$.
    \item Change gamma matrices to Euclidean space: $\gamma^\mu \to i\gamma^\mu_E$.
\end{enumerate}
From now on, it is implicit that we are working in Euclidean space, so we do not include the $E$ subscript hereafter.

In the matching, we use a hard-region expansion, defined by $Q^2, (\pi T)^2 \gg  P^2 \sim m^2$ with $Q$ ($P$) the loop (external) momentum, of loop integrals in the (Euclidean) 4D theory. 

To achieve the desired order in external momenta $P$, we iterate the following algebraic identity:
\begin{equation}
 \frac{1}{(Q+P)^2} = \frac{1}{Q^2}\left[1-\frac{P^2+2Q\cdot P}{(Q+P)^2}\right]\,.
\end{equation}
For example, we compute up to $\mathcal{O}(P^6)$ for $\beta_{81}$ while $\mathcal{O}(P^2)$ for $K_3$. In all cases, we neglect $m^2/T^2$ and higher orders in our expansion, as these are naturally much smaller than the leading contribution to every operator in the matching.

With the usual definition for the sum-integral, i.e.
\begin{equation}
    \sumint \equiv T \sum_{n=-\infty}^\infty \int \frac{dq^3}{(2 \pi)^3}\,,
\end{equation}
where $q$ are the components of the 3D momenta and $n$ are the Matsubara modes running in the loop, we express all our results in terms of the following 1-loop bosonic master sum-integral in dimensional regularization: 
\begin{align*}
    I_\alpha^{\beta\gamma} (d) \equiv \sumint^{~~~~\prime} \frac{\left(Q_0^2\right)^\beta \left( \mathbf{q}^2 \right)^\gamma}{\left( Q^2 \right)^\alpha} = &\left(\frac{e^{\gamma_E} \mu^2}{4 \pi}\right)^\epsilon \frac{2T (2 \pi T)^{d-2\alpha+2\beta+2\gamma}}{(4\pi)^{d/2}} \\ &\frac{\Gamma(d/2 + \gamma) \Gamma(-d/2 + \alpha - \gamma)}{\Gamma(d/2) \Gamma(\alpha)} \zeta(-d+2\alpha-2\beta-2\gamma) \stepcounter{equation}\tag{\theequation}\label{eq:master}\,.
\end{align*}
The prime denotes that we are summing over all non-zero thermal modes, as the zero mode contribution cancels with loop contributions to the same diagrams in the 3D EFT when matching. Practically, however, $\sumint^{~\prime} \text{``=''} \sumint$ at one-loop, since the scalar zero mode contribution to the loop integrals in the hard-region expansion will always be a scaleless integral, which vanishes in dimensional regularization.

The master 1-loop fermionic sum-integral is related to the previous one through
\begin{equation}
    \hat{I}_\alpha^{\beta\gamma} (d) = (2^{2 \alpha - 2 \beta - 2 \gamma - d} - 1)~I_\alpha^{\beta\gamma} (d)\,.
\end{equation}
\noindent As we will regularize the sum-integrals using dimensional regularization, we shall set $d=3-2\epsilon$ in all sum-integrals and skip the $d$ argument.

Let us compute, as an example, the hard-region contribution of the cubic scalar term in Eq.~\eqref{eq:LagUV} to the 2-point function up to $\mathcal{O}\left( P^6 \right)$:

\begin{align*}
    \begin{gathered}
    \includegraphics[width=2.1cm]{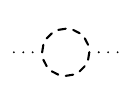}
    \end{gathered}
    &= \frac{(-3! \kappa)^2}{2} \sumint^{~~~~\prime} \frac{1}{Q^2 + m^2} \frac{1}{(Q+P)^2 + m^2} \simeq 18 \kappa^2 \sumint^{~~~~\prime} \frac{1}{Q^2(Q+P)^2} \\
    %
    %
    &\simeq 18 \kappa^2 \sumint^{~~~~\prime} \frac{1}{Q^4} \left[ 1 - \frac{P^2}{Q^2} + \frac{P^4}{Q^4} + \frac{4 \left(P \cdot Q\right)^2}{Q^4} - \frac{P^6}{Q^6} -\frac{12 P^2 \left(P \cdot Q\right)^2}{Q^6} \right. \\
    & \hphantom{\kappa^2 \simeq \sumint^{~~~~\prime} \frac{1}{Q^4} \left[\right.} \left. 
    + \frac{24 P^4 \left(P \cdot Q\right)^2}{Q^8} + \frac{16 \left(P \cdot Q\right)^4}{Q^8} -\frac{80 P^2 \left(P \cdot Q\right)^4}{Q^{10}} + \frac{64 \left(P \cdot Q\right)^6}{Q^{12}} \right] \stepcounter{equation}\tag{\theequation}
\end{align*}
where we have removed all integrals odd in $Q$. Now, we apply the following tensor reduction formulae to the product of momenta:
\begin{align}
    & Q^i Q^j = g^{ij} \frac{Q^2}{d}\,, \\
    & Q^i Q^j Q^k Q^l = g^{ijkl} \frac{Q^4}{d^2 + 2 d}\,, \\
    & Q^i Q^j Q^k Q^l Q^m Q^n = g^{ijklmn} \frac{Q^6}{d^3 + 6d^2 + 8d}\,,
\end{align}
where $g^{i_1 i_2 \dots i_n}$ is the totally symmetric combination of (3-dimensional Euclidean) metric tensors. Finally, we rewrite the sum integral in terms of the master sum-integrals in Eq. \eqref{eq:master}, obtaining
\begin{align*}
    \begin{gathered}
    \includegraphics[width=2.1cm]{FeynmanDiagrams/bead-scalar.pdf}
    \end{gathered}
    = &18 \kappa^2 \left[ I_2^{00} + \frac{1}{3} P^2 \left( I_3^{00} - 4 I_4^{10} \right) + \frac{1}{5} P^4 \left( I_4^{00} - 12 I_5^{10} + 16 I_6^{20} \right) \right.\\
    & \hphantom{18 \kappa^2 \left[\right.} \left. + \frac{1}{7} P^6 \left( I_5^{00} - 24 I_6^{10} + 80 I_7^{20} - 64 I_8^{30} \right) \right]\,. \stepcounter{equation}\tag{\theequation}
\end{align*}

The corresponding amplitude in the 3D EFT reads:
\begin{align*}
    \begin{gathered}
    \includegraphics[width=2.1cm]{FeynmanDiagrams/bead-scalar.pdf}
    \end{gathered}
    = - m_3^2 - K_3 P^2 + 2 \beta_{61} P^4 + 2 P^6 \beta_{81}\,, \stepcounter{equation}\tag{\theequation}
\end{align*}
from where it is straightforward to derive the corresponding matching corrections
for $K_3$, $m_3^2$, $\beta_{61}$ and $\beta_{81}:$
\begin{gather}
    K_3 = - \frac{18}{3} \kappa^2 \left( \bar I_3^{00} - 4 \bar I_4^{10} \right)\,, \quad m_3^2 = -18 \kappa^2 \bar I_2^{00}\,, \\
    \beta_{61} = \frac{9}{5} \kappa^2 \left( \bar I_4^{00} - 12 \bar I_5^{10} + 16 \bar I_6^{20} \right)\,, \quad \beta_{81} = \frac{9}{7} \kappa^2 \left( \bar I_5^{00} - 24 \bar I_6^{10} + 80 \bar I_7^{20} - 64 \bar I_8^{30} \right)\,.
\end{gather}
The $\bar I_{\alpha}^{\beta \gamma}$ denote the non-divergent part of the sum-integral $I_{\alpha}^{\beta \gamma}$ in the $\rm \overline{MS}$ subtraction scheme. Note how, since $I_2^{00}$ is purely divergent, the cubic term contribution to $m_3^2$ does not appear in the complete matching in Eq. \eqref{eq:dim4 matching w scalars}.

\subsection{Matching}

A more general 3D EFT Lagrangian for the scalar zero mode $\varphi$ than the one presented in Eq. \eqref{eq:3deft} also includes the following $\mathbb{Z}_2$-breaking effective operators:
\begin{align}\label{eq:3deft_full}
    &\mathcal{L}_3^{(d=5)} = \alpha_{51} \varphi^5 + \beta_{51} \varphi^2 \partial^2 \varphi\,, \nonumber\\
    &\mathcal{L}_3^{(d=7)} = \alpha_{71} \varphi^7 + \beta_{71} \varphi^4 \partial^2 \varphi + \beta_{72} \varphi \left( \partial^2 \varphi \right)^2 + \beta_{73} \left(\partial \phi\right)^2 \partial^2 \varphi\,. \nonumber
\end{align}

We now present the matching conditions obtained by equating the non-divergent part of the 1PI diagrams in the UV theory up to one-loop and the tree-level in the complete 3D theory. We choose an appropriate matching scale $\mu = \pi T e^{-\gamma_E}$,
where $\gamma_E$ is the Euler-Mascheroni constant, so that all logarithms vanish. Furthermore, since $[\phi] = [\sqrt{T} \varphi] = 1$, we include the appropriate factors of $T$ to match the energy dimensions between the amplitudes in the matching.

\begin{figure*}[t!]
    \centering
    \begin{subfigure}[t]{0.2\textwidth}
        \centering
            \includegraphics[width=2.5cm]{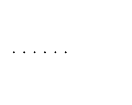}
    \end{subfigure}%
    ~         
    \begin{subfigure}[t]{0.2\textwidth}
        \centering
            \includegraphics[width=2.5cm]{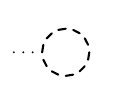}
    \end{subfigure}%
    ~     
    \begin{subfigure}[t]{0.2\textwidth}
        \centering
           \includegraphics[width=2.5cm]{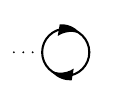}
    \end{subfigure}%
    \vspace{-10pt}
    \caption{One-point 1PI diagrams up to one-loop order in the UV theory \eqref{eq:LagUV}.}
    \label{fig:1-point}
\end{figure*}

\begin{figure*}[t!]
    \centering
    \begin{subfigure}[t]{0.2\textwidth}
        \centering
            \includegraphics[width=2.5cm]{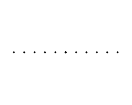}
    \end{subfigure}%
    ~     
    \begin{subfigure}[t]{0.2\textwidth}
        \centering
           \includegraphics[width=2.5cm]{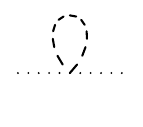}
    \end{subfigure}%
    ~ 
    \begin{subfigure}[t]{0.2\textwidth}
        \centering
           \includegraphics[width=2.5cm]{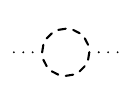}
    \end{subfigure}%
    ~ 
    \begin{subfigure}[t]{0.2\textwidth}
        \centering
           \includegraphics[width=2.5cm]{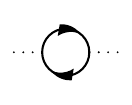}
    \end{subfigure}
    \vspace{-10pt}
    \caption{Two-point 1PI diagrams up to one-loop order in the UV theory \eqref{eq:LagUV}.}
    \label{fig:2-point}
\end{figure*}

\begin{figure*}[t!]
    \centering
    \begin{subfigure}[t]{0.15\textwidth}
        \centering
            \includegraphics[width=2cm]{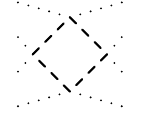}
    \end{subfigure}%
    ~ 
    \begin{subfigure}[t]{0.15\textwidth}
        \centering
           \includegraphics[width=2cm]{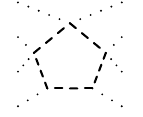}
    \end{subfigure}%
    ~
    \begin{subfigure}[t]{0.15\textwidth}
        \centering
            \includegraphics[width=2cm]{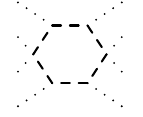}
    \end{subfigure}%
    ~ 
    \begin{subfigure}[t]{0.15\textwidth}
        \centering
         \includegraphics[width=2cm]{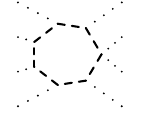}
    \end{subfigure}%
    ~ 
    \begin{subfigure}[t]{0.15\textwidth}
        \centering
          \includegraphics[width=2cm]{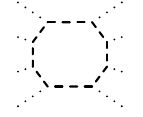}
    \end{subfigure}%
    ~ 
    \begin{subfigure}[t]{0.15\textwidth}
        \centering
          \includegraphics[width=2cm]{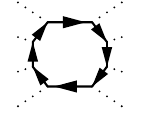}
    \end{subfigure}%
    \caption{Eight-point 1PI diagrams up to one-loop order in the UV theory \eqref{eq:LagUV}.}
    \label{fig:8-point}    
\end{figure*}

In Figs.~\ref{fig:1-point}--\ref{fig:8-point}, we show the representative topologies of the 1PI diagrams in the UV theory. The dotted line represents the zero mode of the scalar, the dashed line represents all scalar modes and the solid line, all fermion modes. We compute the corresponding $n$-point functions off-shell, $\Gamma^{(n)}_\text{UV}$ (1-loop) and $\Gamma^{(n)}_\text{IR}$ (tree-level), using \texttt{FeynArts}~\cite{Hahn:2001} and \texttt{FeynCalc}~\cite{Shtabovenko:2023feyncalc}. 
We assume the following power counting~\cite{Gould:2023jbz}: $P^2/T^2 \sim m^2/T^2 \sim \kappa/T \sim \lambda \sim g^2$, 
and calculate the aforementioned diagrams to $\mathcal{O}(g^8)$.

We match $n$-point functions in powers of external momenta and obtain the matching relations below, which reproduce Eqs.~\eqref{eq:dim4 matching}--\eqref{eq:dim8 matching}
in the limit of vanishing $\kappa$ and $\lambda$:
\begin{equation}
 \sigma_3 = \frac{1}{4} \kappa T^{3/2}\,;
\end{equation}
\begin{equation}
 K_3 = 1 + \frac{g^2}{12\pi^2} + \frac{3 \zeta(3) \kappa^2}{64 \pi^4 T^2} \,,\quad m_3^2 = m^2 + \frac{g^2 T^2}{6} + \lambda T^2\,,\quad \kappa_3 = \kappa\sqrt{T}\,,\quad \lambda_3 = \lambda T\,;
 \label{eq:dim4 matching w scalars}
\end{equation}
\begin{gather}
 \alpha_{51} = - \frac{81 \kappa ^3 \lambda  \zeta (5)}{64 \pi ^6 T^{5/2}} + \frac{27 \kappa  \lambda ^2 \zeta (3)}{8 \pi ^4 \sqrt{T}}\,,\nonumber\\[0.2cm]
 \beta_{51} = \frac{27 \kappa ^3 \zeta (5)}{1024 \pi ^6 T^{7/2}} - \frac{3 \kappa  \lambda  \zeta (3)}{32 \pi ^4 T^{3/2}}\,;
\end{gather}
\begin{gather}
 \alpha_{61} = - \frac{7 g^6 \zeta(3)}{192 \pi^4} + \frac{1215 \kappa^4 \lambda \zeta(7)}{1024 \pi^8 T^4} - \frac{243 \kappa^2 \lambda^2 \zeta(5)}{64 \pi^6 T^2} + \frac{9 \lambda^3 \zeta(3)}{4 \pi^4}\,,\nonumber\\[0.2cm]
 \beta_{61} = - \frac{7 g^2 \zeta(3)}{384 \pi^4 T^2} - \frac{9 \kappa^2 \zeta(5)}{10240 \pi^6 T^4}\,,\nonumber\\[0.2cm]
 \beta_{62} = \frac{35 g^4 \zeta(3)}{576 \pi^4 T} - \frac{135 \kappa^4 \zeta (7)}{4096 \pi^8 T^5}+\frac{45 \kappa^2 \lambda  \zeta (5)}{256 \pi^6 T^3} - \frac{\lambda^2 \zeta(3)}{8 \pi^4 T}\,;
\end{gather}
\begin{gather}
 \alpha_{71} = - \frac{81 \kappa  \lambda ^3 \zeta (5)}{16 \pi ^6 T^{3/2}}\,,\nonumber\\[0.2cm]
 \beta_{71} = - \frac{1215 \kappa ^3 \lambda  \zeta (7)}{4096 \pi ^8 T^{9/2}} + \frac{27 \kappa  \lambda ^2 \zeta (5)}{64 \pi ^6 T^{5/2}}\,,\nonumber\\[0.2cm]
 \beta_{72} = - \frac{9 \kappa  \lambda  \zeta (5)}{1280 \pi ^6 T^{7/2}} + \frac{27 \kappa ^3 \zeta (7)}{8192 \pi ^8 T^{11/2}}\,,\nonumber\\[0.2cm]
 \beta_{73} = - \frac{9 \kappa  \lambda  \zeta (5)}{1280 \pi ^6 T^{7/2}} + \frac{9 \kappa ^3 \zeta (7)}{4096 \pi ^8 T^{11/2}}\,;
\end{gather}
\begin{gather}
 \alpha_{81} = \frac{31 g^8 \zeta (5)}{2048 \pi ^6 T}\,,\nonumber\\[0.2cm]
 \alpha_{82} = - \frac{31 g^4 \zeta (5)}{10240 \pi ^6 T^3} - \frac{189 \kappa ^4 \zeta (9)}{131072 \pi ^{10} T^7} + \frac{9 \kappa ^2 \lambda  \zeta (7)}{1024 \pi ^8 T^5} - \frac{9 \lambda ^2 \zeta (5)}{640 \pi ^6 T^3}\,, \nonumber\\[0.2cm]
 \beta_{81} = - \frac{31 g^2 \zeta(5)}{10240 \pi^6 T^4} - \frac{9 \kappa^2 \zeta (7)}{229376 \pi^8 T^6}\,,\nonumber\\[0.2cm]
 \beta_{82} = \frac{217 g^4 \zeta (5)}{20480 \pi ^6 T^3} - \frac{315 \kappa ^4 \zeta (9)}{262144 \pi ^{10} T^7} + \frac{9 \kappa ^2 \lambda  \zeta (7)}{2048 \pi ^8 T^5} + \frac{3 \lambda ^2 \zeta (5)}{1280 \pi ^6 T^3}\,, \nonumber\\[0.2cm]
 \beta_{83} = \frac{279 g^4 \zeta (5)}{20480 \pi ^6 T^3} - \frac{945 \kappa ^4 \zeta (9)}{262144 \pi ^{10} T^7} + \frac{45 \kappa ^2 \lambda  \zeta (7)}{4096 \pi ^8 T^5} - \frac{9 \lambda ^2 \zeta (5)}{1280 \pi ^6 T^3}\,, \nonumber\\[0.2cm]
 \beta_{84} = -\frac{217 g^6 \zeta (5)}{5120 \pi ^6 T^2} - \frac{1053 \kappa ^2 \lambda ^2 \zeta (7)}{1024 \pi ^8 T^4} + \frac{27 \lambda ^3 \zeta (5)}{80 \pi ^6 T^2}\,.\\\nonumber
\end{gather}

We do not present here the amplitudes in each theory for the sake of readability, as the UV contain tens of terms when expanding to higher orders in external momenta.

We would now like to have a measure of how relevant scalar loops are compared to fermion loops in the matching.
Since loops in the EFT Lagrangian with ($\mathcal{L}_3^{\phi\psi}$) and without ($\mathcal{L}_3^{\psi}$) scalar contributions generate different EFT operators off-shell, we make the comparison in the physical basis in which redundancies are eliminated.

For this purpose, we first remove the tadpole from $\mathcal{L}_3^{\phi\psi}$ through a constant shift of the field $\varphi \to \varphi + a$, where $a$ is perturbatively calculated so the tadpole term vanishes. Next, we canonically normalize the Lagrangian by taking $\varphi \to \varphi/\sqrt{K_3}$. Finally, we perform the appropriate field redefinitions to remove the redundant operators as described in Section~\ref{sec:theory}. An analogous procedure with $\mathcal{L}_3^{\psi}$, in which case the tadpole term is absent, leads to Eqs.~\eqref{eq:initial}--\eqref{eq:final}.

For every physical Wilson coefficient we compute the relative difference $\Delta_i$ of fermion loop contributions $c_i^\psi$ as compared to the joint contribution of fermions and scalars $c_i^{\phi\psi}$, that is:
\begin{equation}
    \Delta_i = \frac{c_i^\psi - c_i^{\phi\psi}}{c_i^{\phi\psi}}\
    \label{eq:delta}
\end{equation}
for sizable values of the Yukawa coupling, $g$.
Since this coefficient controls the relevance of the fermion loops in the matching, we expect that for $g \sim 1$, where strong PTs occur, including scalars will not change the physical coefficients significantly. Indeed, for example, for model B (see Fig.~\ref{fig:tnalphabeta}) with $g=1$ and for a reference temperature of $T=100$ GeV, we 
find that $\Delta_{51} \thickapprox 0.07$, $\Delta_{61} \thickapprox 0.09$, $\Delta_{71} \thickapprox 0.05$, $\Delta_{81} \thickapprox 0.001$ and $\Delta_{82} \thickapprox 0.01$.

\newpage

\begin{figure*}[t!]
    \centering
    \begin{subfigure}[t]{0.2\textwidth}
        \centering
           \includegraphics[width=2.5cm]{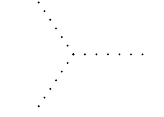}
    \end{subfigure}%
    ~     
    \begin{subfigure}[t]{0.2\textwidth}
        \centering
           \includegraphics[width=2.5cm]{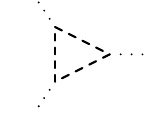}
    \end{subfigure}%
    ~ 
    \begin{subfigure}[t]{0.2\textwidth}
        \centering
            \includegraphics[width=2.5cm]{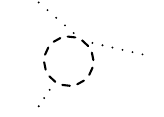}
    \end{subfigure}%
    ~ 
    \begin{subfigure}[t]{0.2\textwidth}
        \centering
            \includegraphics[width=2.5cm]{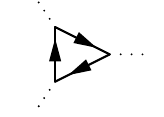}
    \end{subfigure}%
    \caption{Three-point 1PI diagrams up to one-loop order in the UV theory \eqref{eq:LagUV}.}
\end{figure*}

\begin{figure*}[t!]
    \centering
    \begin{subfigure}[t]{0.2\textwidth}
        \centering
           \includegraphics[width=2cm]{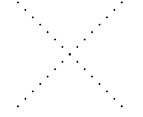}
    \end{subfigure}%
    ~
    \begin{subfigure}[t]{0.2\textwidth}
        \centering
            \includegraphics[width=2cm]{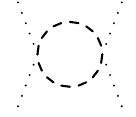}
    \end{subfigure}%
    ~     
    \begin{subfigure}[t]{0.2\textwidth}
        \centering
           \includegraphics[width=2cm]{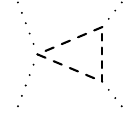}
    \end{subfigure}%
    ~ 
    \begin{subfigure}[t]{0.2\textwidth}
        \centering
          \includegraphics[width=2cm]{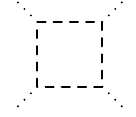}
    \end{subfigure}%
    ~ 
    \begin{subfigure}[t]{0.2\textwidth}
        \centering
           \includegraphics[width=2cm]{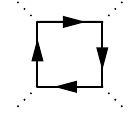}
    \end{subfigure}%
    \caption{Four-point 1PI diagrams up to one-loop order in the UV theory \eqref{eq:LagUV}.}
\end{figure*}

\begin{figure*}[t!]
    \centering
    \begin{subfigure}[t]{0.2\textwidth}
        \centering
           \includegraphics[width=2.5cm]{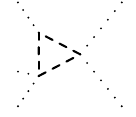}
    \end{subfigure}%
    ~
    \begin{subfigure}[t]{0.2\textwidth}
        \centering
          \includegraphics[width=2.5cm]{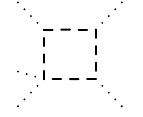}
    \end{subfigure}%
    ~     
    \begin{subfigure}[t]{0.2\textwidth}
        \centering
           \includegraphics[width=2.5cm]{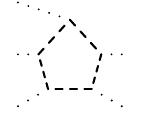}
    \end{subfigure}%
    ~ 
    \begin{subfigure}[t]{0.2\textwidth}
        \centering
            \includegraphics[width=2.5cm]{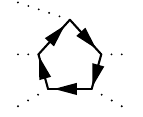}
    \end{subfigure}%
    \caption{Five-point 1PI diagrams up to one-loop order in the UV theory \eqref{eq:LagUV}.}
\end{figure*}

\begin{figure*}[t!]
    \centering
    \begin{subfigure}[t]{0.2\textwidth}
        \centering
           \includegraphics[width=2cm]{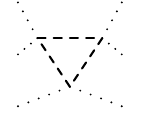}
    \end{subfigure}%
    ~     
    \begin{subfigure}[t]{0.2\textwidth}
        \centering
               \includegraphics[width=2cm]{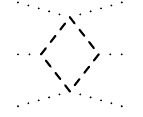}
    \end{subfigure}%
    ~  
    \begin{subfigure}[t]{0.2\textwidth}
        \centering
               \includegraphics[width=2cm]{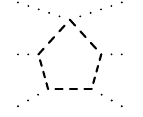}
    \end{subfigure}%
    ~
    \begin{subfigure}[t]{0.2\textwidth}
        \centering
               \includegraphics[width=2cm]{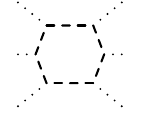}
    \end{subfigure}%
    ~ 
    \begin{subfigure}[t]{0.2\textwidth}
        \centering
              \includegraphics[width=2cm]{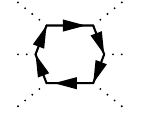}
    \end{subfigure}%
    \caption{Six-point 1PI diagrams up to one-loop order in the UV theory \eqref{eq:LagUV}.}
\end{figure*}

\begin{figure*}[t!]
    \centering
    \begin{subfigure}[t]{0.2\textwidth}
        \centering
           \includegraphics[width=2cm]{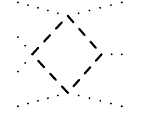}
    \end{subfigure}%
    ~ 
    \begin{subfigure}[t]{0.2\textwidth}
        \centering
           \includegraphics[width=2cm]{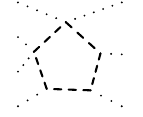}
    \end{subfigure}%
    ~
    \begin{subfigure}[t]{0.2\textwidth}
        \centering
           \includegraphics[width=2cm]{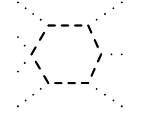}
    \end{subfigure}%
    ~ 
    \begin{subfigure}[t]{0.2\textwidth}
        \centering
          \includegraphics[width=2cm]{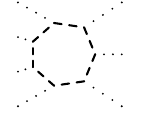}
    \end{subfigure}%
    ~ 
    \begin{subfigure}[t]{0.2\textwidth}
        \centering
            \includegraphics[width=2cm]{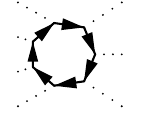}
    \end{subfigure}%
    \caption{Seven-point 1PI diagrams up to one-loop order in the UV theory \eqref{eq:LagUV}.}
\end{figure*}

\section{Perturbative bounce}
\label{app:bounce}

\noindent
We provide here a derivation of the equations defining the perturbative bounce solution and the corresponding bounce action.

\subsection{Equations of motion and bounce action}

Let us now consider the perturbative expansion of a generic functional of the fields $\phi$:
\begin{equation}
    \mathcal{F}[\phi] = 
    \mathcal{F}^{(0)}[\phi]
    + \epsilon \mathcal{F}^{(1)}[\phi]
    + \epsilon^2 \mathcal{F}^{(2)}[\phi]
    + \mathcal{O}(\epsilon^3).
\end{equation}
We will apply the arguments here to two concrete functionals $\mathcal{F}$: the action $S$, and its functional derivative
\begin{equation}
    \mathcal{E}_{x}[\phi] = \left.\frac{\delta S}{\delta \varphi(x)}\right|_{\phi}\,.
\end{equation}
Both of them admit a functional Taylor series expansion of $S$ around any function $\phi$, as
\begin{equation}
    \mathcal{F}[\phi + \eta] = \mathcal{F}[\phi] +
    \int dx \frac{\delta \mathcal{F}}{\delta \phi (x)} \eta(x) + 
    \frac{1}{2} \int dx dy \frac{\delta^2 \mathcal{F}}{\delta\phi(x)\delta\phi(y)} \eta(x) \eta(y) + \mathcal{O}(\eta^3)~,
\end{equation}
for small enough $\eta$,
and where all variational derivatives are evaluated at $\phi$. If we choose $\phi = \varphi_c^{(0)}$ and $\eta = \epsilon \varphi^{(1)}_c + \epsilon^2 \varphi^{(2)}_c + \mathcal{O}(\epsilon^3)$, the expansion becomes
\begin{equation}
    \mathcal{F}[\varphi_c] = 
    \mathcal{F}[\varphi^{(0)}] +
    \epsilon \left.\frac{\delta \mathcal{F}}{\delta \varphi_{x}}\right|_{\varphi^{(0)}} \varphi^{(1)}_{x} +  
    \epsilon^2 
        \left( \left.\frac{\delta \mathcal{F}}{\delta \varphi_{x}}\right|_{\varphi^{(0)}} \varphi^{(2)}_{x} + \frac{1}{2} \left. \frac{\delta^2 \mathcal{F}}{\delta\varphi_{x}\delta\varphi_{y}} \right|_{\varphi^{(0)}} \varphi^{(1)}_{x} \varphi^{(1)}_{y} \right) + \mathcal{O}(\epsilon^3) ~ ,
\end{equation}
where we have used the shorthand notation $\frac{\delta \mathcal{F}}{\delta \varphi_x} f_x \equiv \int dx \frac{\delta \mathcal{F}}{\delta \varphi(x)} f(x)$. For the sake of brevity, we have also omitted the subindex $c$ in $\varphi_c^{(i)}$, and all variational derivatives are understood to be evaluated at $\varphi_c^{(0)}$. We use these conventions for the rest of this appendix.
Now, using the expansion of $\mathcal{F}$ itself in $\epsilon$, the final perturbative expansion of $\mathcal{F}$ evaluated at $\varphi_c$ reads:
\begin{multline}
     \mathcal{F}[\varphi_c] = ~ 
    \mathcal{F}^{(0)}[\varphi^{(0)}] 
    + \epsilon 
        \left( \mathcal{F}^{(1)}[\varphi^{(0)}] + \frac{\delta \mathcal{F}^{(0)}}{\delta \varphi_x} \varphi^{(1)}_x \right)
      \\ 
     + \epsilon^2 
        \left( 
        \mathcal{F}^{(2)}[\varphi^{(0)}] + 
        \frac{\delta \mathcal{F}^{(0)}}{\delta \varphi_x} \varphi^{(2)}_x + 
        \frac{\delta \mathcal{F}^{(1)}}{\delta \varphi_x} \varphi^{(1)}_x + 
        \frac{1}{2} \frac{\delta^2 \mathcal{F}^{(0)}}{\delta\varphi_x\delta\varphi_y}  \varphi^{(1)}_x \varphi^{(1)}_y \right) 
         + \mathcal{O}(\epsilon^3)\,.
    \label{eq:perturbative expansion S}
\end{multline}

Since $\varphi_c$ is an extremal of $S$, we have that $\mathcal{E}_x[\varphi_c] = 0$ for all $x$. Thus, setting $\mathcal{F} = \mathcal{E}_x$ in Eq.~\eqref{eq:perturbative expansion S}, one obtains that the coefficient of each power of $\epsilon$ must vanish. This leads to a set of functional equations that fix, at each order $\epsilon^i$, the function $\varphi^{(i)}$.
From the zeroth-order coefficient, we read that $\varphi^{(0)}$ is actually an extremal of $S^{(0)}$:
\begin{equation}
    \mathcal{E}_x^{(0)}[\varphi^{(0)}] =  \left.\frac{\delta S^{(0)}}{\delta \varphi(x)}\right|_{\varphi^{(0)}} = 0 ~.
    \label{eq:diffeq for phi0}
\end{equation}
Likewise, from the first order term, we obtain:
\begin{equation}
    \mathcal{E}_x^{(1)}[\varphi^{(0)}] +  \frac{\delta \mathcal{E}_x^{(0)}}{\delta \varphi_y} \varphi^{(1)}_y = 0
    \implies
    \left.\frac{\delta S^{(1)}}{\delta \varphi(x)}\right|_{\varphi^{(0)}} +
   \int dy~  \left.\frac{\delta^2 S^{(0)}}{\delta \varphi(y) \delta \varphi(x)}\right|_{\varphi^{(0)}}\varphi^{(1)}(y) = 0 ~.
    \label{eq:diffeq for phi1}
\end{equation}
These functional equations can be transformed into ordinary differential equations if the action is local, i.e.
\begin{equation}
    S^{(i)}[\varphi]=\int dx ~ \mathcal{L}^{(i)}(\varphi, \partial_\mu \varphi, \partial_\mu \partial_\nu \varphi,\dots)\,.
\end{equation}
In this case, both Eq.~\eqref{eq:diffeq for phi0} and the first term of Eq.~\eqref{eq:diffeq for phi1}, corresponding to the first functional derivative of $S^{(0)}$ and $S^{(1)}$, can be handled just applying the generalized Euler-Lagrange equations:
\begin{equation}
   \frac{\delta S^{(i)}}{\delta \varphi}
   =
   \frac{\partial \mathcal{L}^{(i)}}{\partial\varphi} - \partial_\mu \left(\frac{\partial \mathcal{L}^{(i)}}{\partial(\partial_\mu \varphi)}\right)
    + \partial_\mu \partial_\nu \left(\frac{\partial \mathcal{L}^{(i)}}{\partial(\partial_\mu \partial_\nu \varphi)}\right) + \dots ~ .
    \label{eq:generalized_Euler-Lagrange}
\end{equation}
Notice that Eq.~\eqref{eq:generalized_Euler-Lagrange}, when evaluated at $\varphi^{(0)}$, is in general non-vanishing because $\varphi^{(0)}$ is a solution of the EOM for $S^{(0)}$, and not for $S^{(i)}$ with $i\neq 0$.

The second functional derivative (which appears in the second term of Eq.~\eqref{eq:diffeq for phi1}), can be computed as the coefficient of $\eta(x) \eta(y) / 2$ in the expansion of  a local $S[\varphi(x) + \varepsilon \eta(x)]$ in powers of $\varepsilon$, with $\eta(x)$ being a generic function. By doing so, and taking into account that the second functional derivative appears under a double integral (e.g., in $x$ and $y$), it is possible to find the following formal expression:
\begin{multline}
    \frac{\delta^2 S^{(0)}}{\delta \varphi(x) \delta \varphi(y)} =
    \left[ \frac{\partial^2 \mathcal{L}^{(0)}}{\partial \varphi^2} - \partial_\mu \left(\frac{\partial^2 \mathcal{L}^{(0)}}{\partial \varphi \partial(\partial_\mu \varphi)} \right) \right] \delta_{xy}
    \\
    - \partial_\mu \left(\frac{\partial^2 \mathcal{L}^{(0)}}{\partial(\partial_\mu \varphi) \partial(\partial_\nu \varphi)} \right) \frac{\partial}{\partial y^\nu} (\delta_{xy})
    -  \left(\frac{\partial^2 \mathcal{L}^{(0)}}{\partial(\partial_\mu \varphi) \partial(\partial_\nu \varphi)} \right) \frac{\partial^2 }{\partial x^\mu \partial y^\nu} (\delta_{xy})\,,
\end{multline}
where $\delta_{xy}\equiv \delta(x-y)$ is the Dirac delta and we have supposed that $\mathcal{L}^{(0)}$ only depends on the field and its first derivative. When multiplied by another function $f(y)$ and integrated over $y$, we can make use of integration by parts to leave no derivatives acting on $\delta_{xy}$, thus obtaining first and second derivatives of $f(y)$. Finally, the $\delta_{xy}$ cancels out the integral over $y$; hence all functions are evaluated on a single variable $x$. This way, Eq.~\eqref{eq:diffeq for phi1} becomes a second order differential equation for $\varphi^{(1)}(x)$.

We now switch to the case $\mathcal{F} = S$.
Altogether, using Eqs.~\eqref{eq:diffeq for phi0} and \eqref{eq:diffeq for phi1}, with the appropriate transformations just described, into the perturbative expansion for $S$ in Eq.~\eqref{eq:perturbative expansion S}, we find a simplified expression for $S[\varphi_c]$:
\begin{equation}
     S[\varphi_c] = ~ 
    S^{(0)}[\varphi^{(0)}] 
    + \epsilon ~ S^{(1)}[\varphi^{(0)}] 
     + \epsilon^2 
        \left( 
        S^{(2)}[\varphi^{(0)}] +  
        \frac{1}{2} \frac{\delta S^{(1)}}{\delta \varphi_x} \varphi^{(1)}_x 
         \right) 
         + \mathcal{O}(\epsilon^3)\,.
    \label{eq:perturbative expansion S (reduced form)}
\end{equation}

In order to clarify the use of these equations, we show explicitly how to get from Eq.~\eqref{eq:diffeq for phi1} to Eq.~\eqref{eq:diff equation for phi1 model}. First, since $\varphi$ has radial symmetry, the action can be written as $S=4\pi\int dr \, r^2 \mathcal{L}$. Thus, the second term in Eq.~\eqref{eq:diffeq for phi1} can be written as follows:
\begin{multline}
      \int dr' \left.\frac{\delta^2 S^{(0)}}{\delta \varphi(r) \delta \varphi(r')} \right|_{\varphi^{(0)}} \varphi^{(1)}(r') = 
      4 \pi r^2 \left[
      \left( 
      \frac{\partial^2 \mathcal{L}^{(0)}}{\partial \varphi^2} - \partial_r \frac{\partial^2 \mathcal{L}^{(0)}}{\partial \varphi \partial \dot\varphi} - \frac{2}{r} \frac{\partial^2 \mathcal{L}^{(0)}}{\partial \varphi \partial \dot\varphi}
      \right) \varphi^{(1)}(r) \right. \\
      \left.
     - \left(
\partial_r \frac{\partial^2 \mathcal{L}^{(0)}}{\partial \dot\varphi^2} + \frac{2}{r} \frac{\partial^2 \mathcal{L}^{(0)}}{\partial \dot\varphi^2}
      \right) \dot \varphi^{(1)}(r) -
      \left(
      \frac{\partial^2 \mathcal{L}^{(0)}}{\partial \dot\varphi^2} 
      \right) \ddot \varphi^{(1)}(r)
      \right] ~ ,
\end{multline}
where we have used integration-by-parts to remove derivatives of $\delta_{rr'}$.
Also, we use $\dot \varphi$ to represent the derivative of $\varphi$ with respect to $r$. 
If $\mathcal{L}^{(0)}$ has only a kinetic term and a potential, i.e., $\mathcal{L}^{(0)} = \frac{1}{2} (\partial_r \varphi)^2 + V^{(0)}(\varphi)$, we finally get:
\begin{equation}
    \int dr' \left.\frac{\delta^2 S^{(0)}}{\delta \varphi(r) \delta \varphi(r')} \right|_{\varphi^{(0)}} \varphi^{(1)}(r') = 
    4 \pi r^2 \left[
    V^{(0)''}(\varphi^{(0)}) ~ \varphi^{(1)}(r) - \frac{2}{r} \dot\varphi^{(1)}(r) - \ddot\varphi^{(1)}(r)
    \right] ~,
\end{equation}
from which we immediately obtain Eq.~\eqref{eq:diff equation for phi1 model} upon inserting this into Eq.~\eqref{eq:diffeq for phi1}.

\subsection{Boundary conditions}
In order to fully characterise the perturbative bounce, we need to discuss the (perturbative) boundary conditions. Generally, we have that  $\dot \varphi_c(0) = 0$ and $\lim_{r\to\infty}\varphi_c(r) = \varphi_F$ with $\varphi_c(0) \neq \varphi_F$.
From the first condition, we obtain:
\begin{equation}
     0 = \dot \varphi_c(0) = \dot\varphi^{(0)}(0) + \epsilon \dot\varphi^{(1)}(0) + \epsilon^2 \dot\varphi^{(2)}(0) + \dots ~ ,
\end{equation}
Given that $\epsilon$ is arbitrary, it must be the case that $\dot \varphi^{(n)}(0) = 0$ for every $n \in \mathbb{N}$. The second condition is the statement that, as $r\to \infty$, the value of $\varphi_c$ approaches the false minimum of the potential $V$. Essentially, the only thing we need to do is to minimise the potential in a consistent perturbative way. The process is very similar to the one described earlier, but applied to a plain function $V$ instead of a functional $S$.
Indeed, let 
\begin{equation}
    V=V^{(0)} + \epsilon V^{(1)} + \epsilon^2 V^{(2)} + \dots ~ ,
    \qquad 
    \lim_{r\to\infty}\varphi_c(r) \equiv \varphi_{\infty} = \varphi_\infty^{(0)} + \epsilon \varphi_\infty^{(1)} + \epsilon^2 \varphi_\infty^{(2)} + \dots ~ ,
    \label{eq:potential-expansion-reduced}
\end{equation}
be the potential and the asymptotic value of the solution as $r\to \infty$, both expanded in the perturbative parameter $\epsilon$. Then, Taylor expanding in $V'(\varphi_\infty)$ for $\epsilon\ll 1$ and collecting $\epsilon$ terms, we arrive at:
\begin{multline}
     V'(\varphi_\infty) = ~ 
    V^{(0)'}(\varphi_\infty^{(0)}) 
    + \epsilon 
        \left( V^{(1)'}(\varphi_\infty^{(0)}) +  V^{(0)''}(\varphi_\infty^{(0)}) ~\varphi^{(1)}_\infty \right)
        \\
     + \epsilon^2 
        \left\{
        V^{(2)'}(\varphi_\infty^{(0)}) + 
        V^{(0)''}(\varphi_\infty^{(0)})  ~ \varphi^{(2)}_\infty + 
        V^{(1)''}(\varphi_\infty^{(0)}) ~ \varphi^{(1)}_\infty + 
        \frac{1}{2}  V^{(0)'''}(\varphi_\infty^{(0)}) ~ \left( \varphi^{(1)}_\infty\right)^2 \right\}
         + \mathcal{O}(\epsilon^3) ~.
    \label{eq:perturbative expansion V}
\end{multline}
Since $\varphi_\infty$ minimises $V$, then $V'(\varphi_\infty)=0$ holds, and, being $\epsilon$ an arbitrary parameter, it has to do it order by order in $\epsilon$. For instance, 
\begin{equation}
    V^{(0)'}(\varphi_\infty^{(0)}) = 0 ~,
%
%
    \qquad \text{and} \qquad
    \varphi^{(1)}_\infty  = - \frac{V^{(1)'}(\varphi_\infty^{(0)})}{V^{(0)''}(\varphi_\infty^{(0)})}~ .
\end{equation}
Hence, each term in the perturbative bounce solution is uniquely characterized by the differential equations coming from Eq.~\eqref{eq:perturbative expansion S} upon replacing $S$ with $\delta S/\delta \varphi$, together with the boundary conditions $\dot\varphi^{(i)}(0)=0$, $\lim_{r\to\infty}\varphi^{(i)}(r)=\varphi^{(i)}_\infty$, where $\varphi^{(i)}_\infty$ satisfy the algebraic identities that Eq.~\eqref{eq:perturbative expansion V} entails. Notice that $\varphi^{(0)}$ is actually the bounce solution of $S^{(0)}$.

\subsection{Results to arbitrary order}
In this final section, we deduce a general formula for the coefficient of any given order in the expansion in power of $\epsilon$ of a generic functional around an extremal.
This generalizes both Eq.~\eqref{eq:perturbative expansion S (reduced form)} and Eq.~\eqref{eq:potential-expansion-reduced}.
Let
\begin{equation}
    \mathcal{G}[\varphi] = \sum_{i=0}^\infty \epsilon^i  \mathcal{G}^{(i)}[\varphi]~, \qquad
    \varphi = \sum_{i=0}^\infty \epsilon^i \varphi^{(i)} ~,
\end{equation}
and consider the Taylor expansion
\begin{equation}
    \mathcal{G}[\varphi] = \mathcal{G}[\varphi^{(0)}] + \sum_{n=1}^\infty \frac{1}{n!} \frac{\delta^n \mathcal{G}}{\delta \varphi^n} (\varphi - \varphi^{(0)})^n~,
    \label{eq:series expansion}
\end{equation}
where
\begin{equation}
    \frac{\delta^n \mathcal{G}}{\delta \varphi^i} (\varphi - \varphi^{(0)})^n 
    \equiv
    \int dx_1 \dots dx_n \frac{\delta^n \mathcal{G}}{\delta \varphi(x_1) \dots \delta \varphi(x_n)} \left[\varphi(x_1) - \varphi^{(0)}(x_1)\right] \dots \left[\varphi(x_n) - \varphi^{(0)}(x_n)\right] ~.
\end{equation}
This series expansion is also valid for plain functions instead of functionals, just replacing $\delta$ for $\partial$ in Eq.~\eqref{eq:series expansion}.
Writing $\mathcal{G}_i \equiv \mathcal{G}_i[\varphi^{(0)}]$, we have:
\begin{equation}
    \mathcal{G}[\varphi] = 
    \sum_{i=0}^\infty \epsilon^i \mathcal{G}_i +
    \sum_{n=1}^\infty  \sum_{i=0}^\infty \frac{\epsilon^i}{n!} \frac{\delta^n \mathcal{G}_i}{\delta \varphi^n} \left(\sum_{j=1}^\infty \epsilon^j \varphi^{(j)} \right)^n ~.
    \label{eq:general perturbation 1}
\end{equation}
Now, the expansion of the $n$-th power in the last part of the expression reads:
\begin{equation}
    \left( \sum_{j=1}^\infty \varphi^{(j)} \right)^n
    =
    \sum_{k=n}^\infty \epsilon^k \left( \sum_{\substack{r_1+\dots + r_n=k \\ 1\leq r_1,\dots, r_n \leq k}} \varphi^{(r_1)} \dots \varphi^{(r_n)} \right) ~,
\end{equation}
where $\{r_1,\dots,r_n\}$ is a sequence of $n$ non-negative integers. 

Notice that a permutation of the same sequence does not alter the result inside the sum. So, we can sort the sequence in a canonical way and simply add a factor accounting for the number of possible permutations. This way, the counting of the possibilities for the different $r_1,\dots, r_n$ is drastically reduced.
If we define $\{t_1,\dots, t_m\}$ to be the sequence $r$ after removing duplicate elements and $s_i$ the number of times that $t_i$ appears in $r$, then the number of possible permutations is given by the multinomial coefficient
\begin{equation}
\binom{n}{s_1, \dots, s_m} = \frac{n!}{s_1! \dots s_m!}\,.
\end{equation}
Putting all together, the expression in Eq.~\eqref{eq:general perturbation 1} reads:
\begin{align}
\notag
    \mathcal{G}[\varphi] & = 
    \sum_{i=0}^\infty \epsilon^i \mathcal{G}_i +
    \sum_{n=1}^\infty  \sum_{i=0}^\infty \sum_{k=n}^\infty \frac{\epsilon^{i+k}}{n!} \frac{\delta^n \mathcal{G}_i}{\delta \varphi^n} 
    \sum_{\substack{r_1+\dots + r_n=k \\ 1\leq r_1 \leq \dots \leq r_n \leq k}} 
    \frac{n!}{s_1!\dots s_m!}
    \left(\varphi^{(t_1)}\right)^{s_1} \dots  \left(\varphi^{(t_m)}\right)^{s_m}
    \\
    & = 
    \sum_{i=0}^\infty \epsilon^i \mathcal{G}_i +
    \sum_{n=1}^\infty \sum_{i,k=0}^\infty  \frac{\epsilon^{i+k+n}}{n!} \frac{\delta^n \mathcal{G}_i}{\delta \varphi^n} 
    \sum_{\substack{r_1+\dots + r_n=k+n \\ 1\leq r_1 \leq \dots \leq r_n \leq k+n}} 
    \frac{n!}{s_1!\dots s_m!}
    \left(\varphi^{(t_1)}\right)^{s_1} \dots  \left(\varphi^{(t_m)}\right)^{s_m}\,.
\end{align}
Finally, we can do a renaming of the indices, such that $\alpha=i+k+n$, obtaining:
\begin{equation}
    \mathcal{G}[\varphi] = 
    \sum_{\alpha=0}^\infty \left( \mathcal{G}_\alpha +  \sum_{n=1}^\alpha \sum_{i=0}^{\alpha-n}
    \frac{1}{n!} \frac{\delta^n \mathcal{G}_i}{\delta \varphi^n} 
    \sum_{\substack{r_1+\dots + r_n=\alpha-i \\ 1\leq r_1 \leq \dots \leq r_n \leq \alpha-i}} 
    \frac{n!}{s_1!\dots s_m!}
    \left(\varphi^{(t_1)}\right)^{s_1} \dots  \left(\varphi^{(t_m)}\right)^{s_m}
    \right) \epsilon^\alpha ~ .
    \label{eq:general form of perturbative expansion}
\end{equation}
All perturbative bounce computations can be easily derived from here. Thus, the differential equation for $\varphi^{(n)}$ results from replacing $\mathcal{G}$ by $\frac{\delta S}{\delta \varphi}$ and equating the $n$-th order of $\epsilon$ to zero. Likewise,  the boundary condition as $r\to\infty$ for $\varphi^{(n)}$ comes from replacing $\mathcal{G}$ by $V'$.

For practical purposes, though, in this work we have implemented a  \texttt{Mathematica} code which automatically computes variational derivatives to any order and gives the right coefficient in $\epsilon$ in the perturbative expansion.

\bibliographystyle{style} 

\bibliography{refs}

\end{document}